\pdfoutput=1
\documentclass[journal,doublecolumn]{IEEEtran}

\ifCLASSINFOpdf
\else
\fi
\usepackage[utf8]{inputenc} 
\usepackage[T1]{fontenc}   
\usepackage{hyperref}      
\usepackage{url}           
\usepackage{booktabs}       
\usepackage{amsfonts}       
\usepackage{nicefrac}       
\usepackage{microtype}      
\usepackage{xcolor}
\usepackage{graphicx} 
\usepackage{tabulary}
\usepackage{xspace}
\usepackage{algorithm}
\usepackage[subtle]{savetrees}
\usepackage{soul}
\usepackage{caption}
\usepackage{subcaption}
\usepackage{algpseudocode}
\usepackage{tabularx}
\usepackage{soul}
\usepackage{multirow} 
\usepackage{amsmath}
\usepackage{makecell}
\usepackage{subcaption}
\newtheorem{theorem}{Theorem}

\newcommand{\descr}[1]{\smallskip \noindent \textbf{#1}}
\newcommand{\descrit}[1]{\smallskip \noindent}
\newcommand{\revised}[1]{{\color{black}{#1}}}
\newcommand{\revisedTwo}[1]{{\color{black}{#1}}}
\newcounter{functionality}
\newenvironment{functionality}[1]
  {\par\addvspace{\topsep}
   \noindent
   \tabularx{\linewidth}{ @{} X  @{} }
    \hline
    \refstepcounter{functionality}\textbf{Functionality \thefunctionality} #1 \\
    \hline
    }
  { \\
   \hline
   \endtabularx
   \par\addvspace{\topsep}}

\newcommand{\sys}{\textsc{CURE}\xspace}
\newif\ifcomment
\commenttrue 
\ifcomment

	\newcommand{\sinem}[1]{\textbf{\em\color{red}[SS: #1]}}

\else  
    \newcommand\sinem[1]{}
\fi
\begin{document}
%
\title{\sys: Privacy-Preserving Split Learning Done Right}
\author{
    \IEEEauthorblockN{Halil Ibrahim Kanpak\IEEEauthorrefmark{1}, Aqsa Shabbir\IEEEauthorrefmark{2}, Esra Genç\IEEEauthorrefmark{2}, Alptekin Küpçü\IEEEauthorrefmark{1}\IEEEauthorrefmark{3}, Sinem Sav\IEEEauthorrefmark{2}\IEEEauthorrefmark{3}\\ }
    \IEEEauthorblockA{\IEEEauthorrefmark{1}Koç University, Istanbul, Turkey\\
    }
     \IEEEauthorblockA{\IEEEauthorrefmark{2}Bilkent University, Ankara, Turkey\\
    }
    \IEEEauthorblockA{\IEEEauthorrefmark{3}\textbf{Correspondence:} akupcu@ku.edu.tr and sinem.sav@cs.bilkent.edu.tr \\}

}

\maketitle
\begin{abstract}
Training deep neural networks often needs large datasets stored and processed in the cloud, and in sensitive fields like healthcare, these workflows must follow strict privacy rules. Split Learning (SL), a framework that divides model layers between client(s) and server(s), is widely adopted for distributed model training. While SL reduces privacy risks by limiting server access to the full parameter set, previous research has identified that intermediate outputs exchanged between server and client can compromise the client's data privacy. Homomorphic encryption (HE)-based solutions exist, but they often impose prohibitive computational burdens. To address these challenges, we propose CURE, a novel system based on HE for the single-client setting that encrypts only the server side of the model and optionally the data. CURE enables secure SL while substantially improving communication and parallelization. We propose packing schemes for efficient execution of deep learning algorithms and generalize them to MLPs and convolutional models, enabling the evaluation of large architectures using our implementations, such as ResNet blocks. We demonstrate that CURE can achieve similar accuracy to plaintext SL, while being up to 210x more efficient in terms of the runtime compared to the state-of-the-art privacy-preserving alternatives. Finally, we propose a novel estimator that enables efficient use of HE in SL settings by recommending an optimal server-client split. 
\end{abstract}
\section{Introduction}

Big data has been a key driver of machine learning (ML) advancements, enabling the training of more complex models. However, massive datasets create storage and processing bottlenecks, making local computation on standard machines impractical. Additionally, handling big data raises privacy concerns due to sensitive information. Collaborative ML enables multiple parties to train a machine learning model without sharing raw data or the model itself. The most popular collaborative ML techniques include federated learning ~\cite{2017federated,konevcny2016federated} and split learning~\cite{Gupta2018}. Federated Learning (FL) enables multiple parties to train a machine learning model without sharing their local data directly. Instead, they share local model updates with a central server, which aggregates these updates to train a global model. Split Learning (SL), on the other hand, splits the neural network (NN) architecture into client-side and server-side models. Thus, it facilitates the training of NNs without sharing the data and/or labels with the server. SL is especially useful in asymmetrical computational resource settings, where clients may lack significant computational power. 

Although FL and SL reduce privacy risks by restricting the server's access to raw data or segments of the model, recent research demonstrates that the client's intermediate model updates, i.e., the gradients shared with the server, can still inadvertently leak the training data or the labels~\cite{pasquini_unleashing_2021,melis2019exploiting,Fu_Ma_Zhu_Hu_Zhao_Jia_Xu_Jin_Zhang_2023,Yu_Wang_Zeng_Zhao_Pang_Wu_2023,mao2023secure,gawron2022feature,hitaj2017deep,ozfatura2024byzantines,parisot2021property,erdougan2022unsplit}. 
Researchers focused on developing new defense strategies to mitigate this leakage in FL using differential privacy (DP)~\cite{shokri2015privacy,abadi2016deep,mcmahan2018LSTM,Wei2020,wu2019value}, homomorphic encryption (HE)~\cite{froelicher2020scalable,poseidon,karakoc2024fault,rhode}, or secure multiparty computation (MPC)~\cite{jayaraman2018distributed,truex2019hybrid,zheng2019helen,Zhang20202,Cock2021,wagh2019securenn,falcon,Zhu2020}. To mitigate various adversarial attacks in SL, several works rely on DP~\cite{thapa2022splitfed,titcombe2021practical,abuadbba2020can,wang2021dplis,yang2022differentially}. However, DP-based learning requires low privacy budgets, resulting in lower accuracy~\cite{rahman2018membership}. Some works employ outlier detection~\cite{erdogan2022splitguard,erdogan2024splitout}. Another line of research employs HE for encrypted training or inference~\cite{Pereteanu_Alansary_Passerat_Palmbach_2022,khan2023split,khan2023love,khan2023more}. While Pereteanu et al. integrate HE for \textit{inference} tasks in SL~\cite{Pereteanu_Alansary_Passerat_Palmbach_2022}, most efforts to improve privacy focus only on U-shaped SL, where the neural network is divided into three segments: the client handles the initial and final layers, while the server processes the intermediate layers~\cite{khan2023split,khan2023love,khan2023more}. This setting assumes that the client holds its own data and labels, necessitating sufficient storage and computational capacity on the client side.

In this work, we focus on privacy-preserving training in SL framework with a single client and a single server, where the server has access to the samples and the client holds the labels. We ensure label confidentiality and, optionally, sample protection. This approach suits scenarios where clients outsource sample storage and parts of the training, such as large-scale genomic datasets. While genomic data, like that related to Autism Spectrum Disorder (ASD), may be stored unencrypted, it typically does not reveal sensitive labels. However, the labels themselves, which are critical for complex traits influenced by numerous genomic variants, remain sensitive and must be protected~\cite{de2014synaptic,satterstrom2020large,norman2019st}.

To ensure data and label confidentiality, we propose \sys, a novel system that leverages HE to encrypt model parameters on the \textit{server side only}. This allows the server to work with an encrypted model, while the client --the data and label owner-- operates on a plaintext model. Encrypting the server-side model allows \sys to mitigate privacy attacks and ensures label privacy by default. Additionally, \sys can optionally encrypt data samples for enhanced data privacy. Thus, our setup protects data and label confidentiality while reducing communication and computation through plaintext training on the client side.

Our contributions can be summarized as follows: (i) We introduce a novel system, \sys, for privacy-preserving SL that ensures the confidentiality of labels and (optionally) the data using HE in single-server, single-client setting. (ii) We propose two packing schemes that ensure efficient computation under various settings, for one-level server operations. (iii) We generalize our packing to support encrypted multi-layer server operations. (iv) We implement a novel estimator that, for the first time, optimally determines the best network split for SL, ensuring efficient use of \sys based on the server's and client's resources. (v) For the first time, we evaluate the performance of ResNet building blocks in an encrypted SL setting. (vi) We develop an open-source API for HE operations in deep learning networks, facilitating the extension of these techniques. Our implementation is available at \url{https://github.com/CRYPTO-KU/CURE-Privacy-Preserving-Split-Learning}.

\section{Related Work}\label{sec:related}
\textbf{Split Learning (SL)}~\cite{Gupta2018} is a machine learning technique that allows model training on distributed datasets without the need to exchange raw data between participants. SL accomplishes this by dividing the ML model into sections, each handled by a different party. It gained recognition with SplitNN~\cite{vepakomma2018split}, 
as a more resource-efficient model than approaches like FL~\cite{shokri2015privacy,2017federated,thapa2021advancements}. It enables various configurations
~\cite{li2023split,satpathy2024collective,poirot2019split,poirot2020split,joshi2022performance}, including the \textit{vanilla configuration}~\cite{vepakomma2018split,majeed2021vanilla,zhang2020split}, where the network is divided at a specific split layer. Additionally, \textit{vertical SL}~\cite{allaart2022vertical,ads2021multi} involves different parties holding different features of the dataset~\cite{navathe1984vertical,navathe1989vertical}. In contrast, \textit{horizontal SL} involves different dataset samples held by various parties to be processed independently~\cite{ceri1982horizontal,ra1993horizontal} to enhance query performance. 
Soon after SL gained recognition, 
several attacks were developed, 
including inference attacks~\cite{pasquini_unleashing_2021,erdougan2022unsplit}, hijacking attacks~\cite{Fu_Ma_Zhu_Hu_Zhao_Jia_Xu_Jin_Zhang_2023}, backdoor attacks~\cite{Yu_Wang_Zeng_Zhao_Pang_Wu_2023,yu2024chronic}, feature distribution attacks~\cite{gawron2022feature}, data reconstruction attacks~\cite{yu2024sia}, and property inference attack~\cite{parisot2021property}. 

\textbf{Privacy-Preserving SL:}
\label{sec:related_ppsplit}
To enhance privacy, several works integrate a mechanism called differential privacy (DP) to SL~\cite{thapa2022splitfed,titcombe2021practical,abuadbba2020can,wang2021dplis,yang2022differentially}. DP adds noise to the data or intermediate values shared between client and server, thereby reducing the accuracy of the results. Our work differs from DP-based methods by combining HE --encryption that supports computation on ciphertext-- with SL, removing this privacy–accuracy tradeoff.

Other works~\cite{Pereteanu_Alansary_Passerat_Palmbach_2022,khan2023love,khan2023split,khan2023more} integrate HE  into SL. By encrypting data or model parameters, any information obtained by attackers is rendered useless without the decryption key. Pereteanu et al.~\cite{Pereteanu_Alansary_Passerat_Palmbach_2022} propose a solution leveraging HE and U-shaped split Convolutional Neural Networks (CNN) to ensure data privacy, specifically designed for fast and secure \textit{inference}. Their model enhances secure \textit{inference} by distributing the model weights between the client and server, with the client computation done in plaintext. 
In contrast, our approach focuses on efficient and secure \textit{training} of the model using advanced packing techniques to optimize communication and computation. By encrypting server-side model parameters and utilizing an \textit{inverted split learning} setup, where the server processes the initial layers and sends intermediate results to the client, we enable collaborative training while preserving confidentiality. 

Khan et al. address the privacy challenge in SL by integrating HE into training to encrypt activation maps before transferring them from the client to the server~\cite{khan2023love,khan2023split,khan2023more}. For this, the authors developed a U-shaped split 1D CNN model, with the initial and final layers are computed by the client and the intermediate layers computed by the server. This design enables clients to protect the privacy of their ground truth labels. In~\cite{khan2023split}, the authors enhanced the model further by ensuring that clients do not need to share either their input training samples or ground truth labels with the server. Similarly, in~\cite{khan2023more}, the authors extended their experiments and introduced batch encryption to optimize memory usage and computational performance when handling encrypted data. 
Nguyen et al.~\cite{nguyen2023split} refined the approach of~\cite{khan2023split} by reducing their privacy leakage and improving communication efficiency utilizing CKKS HE scheme (see Section~\ref{sec:CKKS}). 
Finally, while other approaches integrate HE into SL to enhance privacy, our method stands out by (i) focusing on optimizing training efficiency through HE applied \textit{exclusively to server-side model parameters}, and (ii) implementing an estimator function to optimize the server-client split in constrained settings, improving HE utilization and overall efficiency.
\section{Building Blocks}
\subsection{Neural Networks (NN)}
NN is a model composed of interconnected layers of nodes~\cite{NeuralBook}. During training, the network adjusts connection weights between neurons to minimize the loss between predictions and actual outcomes, using optimization algorithms. Input data (\( X \)) is passed through the network to produce a predicted output (\( \hat{Y} \)). The forward pass predicts \( \hat{Y} \) by applying activation to a linear combination of the layer's weights and the previous layer's activations: $\hat{Y} = \psi(Z_{l}) = \psi(W_{l}O_{l-1} + B_{l})$.
Here, \(l\) is the current layer, \(O_{l-1}\) denotes the output of the previous layer $l-1$, \( \psi \) is the activation function (e.g.,
Sigmoid, ReLU), \(W_{l}\) and \(B_{l}\) denote the weight matrix and the bias vector at layer \(l\), respectively. We denote the linear combination of the weights and the activations as \(Z_{l}\).
After the forward pass, backpropagation~\cite{rumelhart1986learning} updates the network's weights by calculating a loss function (\( J \)) using the predicted output (\( \hat{Y} \)) and the labels (\( Y \)). The gradient of the loss function is calculated by
$g = \frac{\partial J}{\partial Z_{l}} = \frac{\partial J}{\partial \hat{Y}} \cdot \frac{\partial \hat{Y}}{\partial Z_l}$
where \(g\) is the gradient of the loss function with respect to the input \(Z_l\) at layer \(l\).
The parameters (weights and biases) are then updated using:
$ W_{l} \leftarrow W_{l} - \alpha \frac{\partial J}{\partial W_{l}} $ and $B_{l} \leftarrow B_{l} - \alpha \frac{\partial J}{\partial B_{l}}$, where \( \alpha \) is the learning rate.

\subsection{Split Learning (SL)} \label{split_learning}
SL~\cite{vepakomma2018split,poirot2020split,zhang2020split,allaart2022vertical, Gupta2018} is designed to enhance privacy while enabling collaborative model training across multiple entities. In SL, the NN model with a total of $n+k$ layers is divided into two segments: the client-side segment with \( k \) layers and the server-side segment with \( n \) layers. Each client processes its local data (\( X \)) through the first \( k \) layers of the model. The output from the \( k \)-th layer, denoted as \( O_k \), is then transmitted to the server. Instead of transmitting raw data, this intermediate representation is used for subsequent computations. The server then continues the forward pass through the remaining \( n \) layers to compute the predicted output, denoted as \( \hat{Y} \), which is used to evaluate the loss function (\( J \)). The server computes the gradient and sends it back to the client(s). Each client uses this gradient to perform the backpropagation through its \( k \) layers and update its model parameters accordingly. This iterative process continues until the model converges or reaches a predefined number of epochs. For a detailed explanation of our SL architecture and its implementation, see Section~\ref{System Overview}.

SL offers several advantages. First, it enhances privacy by keeping raw data on the client side and sharing only abstract intermediate representations, reducing the privacy risk. Second, SL reduces the client’s computational load, as clients only process \( k \) layers, making it ideal for devices with limited resources. Thus, SL shows promise for achieving secure and efficient collaborative learning across diverse domains. However, adversarial attacks~\cite{pasquini_unleashing_2021,erdougan2022unsplit,Fu_Ma_Zhu_Hu_Zhao_Jia_Xu_Jin_Zhang_2023,Yu_Wang_Zeng_Zhao_Pang_Wu_2023,yu2024chronic,gawron2022feature,liu2024similarity,mao2023secure,hitaj2017deep,parisot2021property,melis2019exploiting,yu2024sia} on SL continue to pose a threat. To address this, we enhance SL by integrating HE, which we detail below. 

\subsection{Homomorphic Encryption (HE)}\label{sec:CKKS}
HE enables computation on ciphertexts, producing encrypted results that, when decrypted, yield the same outcome as if the operations were performed on the plaintext. This is essential for privacy-preserving computations, allowing data to be processed without compromising confidentiality.
In this work, we use the Cheon-Kim-Kim-Song (CKKS)  scheme~\cite{Cheon_Kim_Kim_Song_1970}, which is a leveled HE scheme based on the ring learning with errors (RLWE) problem~\cite{Lyubashevsky}. The scheme is well-suited for supporting approximate (floating-point) precision. CKKS significantly enhances computational efficiency with its \textit{packing capability}, enabling simultaneous processing of multiple data points through Single Instruction, Multiple Data (SIMD) operations on encrypted data. The scheme also has effective noise management strategies, where the noise refers to the small error added to ciphertexts to ensure security. The ring in CKKS is defined as \(\mathbb{Z}[X]/(X^N + 1)\), where \(N\) is a power of two. Key parameters include the cyclotomic ring size (\(N\)), the ciphertext modulus (\(Q\)), the logarithm of the moduli of the ring (\(LogQP\)), the noise parameter ($\sigma$), and the level of the ciphertext (\(L\)) that help manage the depth of the circuit to be evaluated before refreshing the ciphertext through the \(\textsf{Bootstrap}\) operation. The scheme allows packing $N/2$ values to plaintext/ciphertext slots for SIMD operations. The slots of the vector can be rearranged through an operation known as "rotations", which can be computationally expensive. We introduce the key functionalities of the CKKS scheme here:  
\begin{itemize}
    \item \(\textsf{KeyGen}(1^\lambda) \rightarrow \textsf{PK},\textsf{SK}\): Generates a public key (\(\textsf{PK}\)) for encryption and a secret key (\(\textsf{SK}\)) for decryption, given a security parameter (\(\lambda\)) in unary.
    \item \(\textsf{Enc}_{\textsf{PK}}(m) \rightarrow c\): Encrypts a plaintext message (\(m\)) into ciphertext (\(c\)) using \(\textsf{PK}\).
    \item \(\textsf{Dec}_{\textsf{SK}}(c) \rightarrow m\): Decrypts a ciphertext message (\(c\))  back into the plaintext message (\(m\)) using \(\textsf{SK}\).
    \item \(\textsf{Eval}_{\textsf{PK}}(c_v, c_w) \rightarrow c\): Performs arithmetic operations such as addition and multiplication directly on ciphertexts $(c_v,c_w)$, producing a new ciphertext $c$ that represents the result of the operation on the original plaintexts. Each multiplication consumes one level of the ciphertext. 
    \item \(\textsf{Bootstrap}(c) \rightarrow c'\): Refreshes ciphertexts (\(c\)) to produce a fresh ciphertext (\(c'\)) at the initial level. 
\end{itemize}
We denote encrypted ciphertext vectors in bold case, e.g., $\mathbf{X}$, and encoded plaintext vectors in regular case, e.g., $X$, throughout the paper.
We use the notation \(\textsf{Eval}_{\textsf{PK}}^f(\{c_i\}) \rightarrow c\) to denote that a function $f$ is evaluated over a set of ciphertexts $\{c_i\}$ (potentially through multiple basic \textsf{Eval} operations consuming multiple levels of ciphertexts), resulting in the ciphertext $c$.

\section{METHOD}\label{sec:system}

\textbf{Problem Statement.}
We consider an \textit{inverted SL scenario}, where model training is divided between a server and a client. The server potentially has access to the data samples $X$, but not the labels $Y$, which are only known to the client. This setting is motivated by a client who wishes to outsource storage and part of the computation to the server side. Our goal is to enable training within this SL framework while preserving the confidentiality of the labels and, optionally, the data samples. Note that (reconstruction, inference, etc.) attacks on the client side are beyond the scope of this work, as we assume the client owns both the samples and labels but outsources storage and some processing.

\begin{figure}[t]
    \centering
    \subfloat[\revised{Traditional vanilla SL: The client processes data samples $\mathbf{X}$ through $k$ layers, while the server holds the labels $Y$ and computes the remaining $n$ layers. The server then returns the gradients required for updating $W_c$.}]{
        \includegraphics[width=0.85\linewidth]{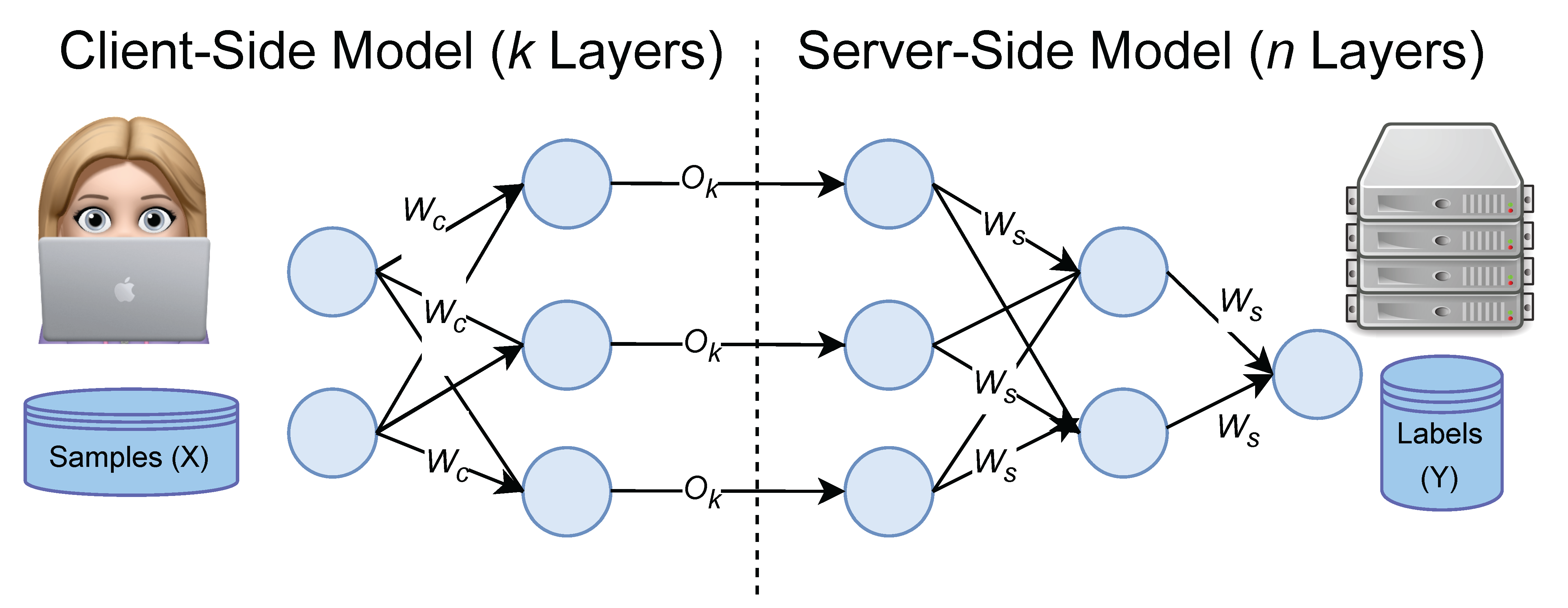}
        \label{fig:vanilla-split}
    }\\[1.3em] 
    \subfloat[\revised{\sys: The inverted SL setup where the server processes data samples $\mathbf{X}$ through $n$ layers, while the client holds the labels $Y$ and processes $k$ layers of the NN. Server-side weights $\mathbf{W}_s$ are encrypted, while client-side weights $W_c$ remain plaintext.}]{
        \includegraphics[width=0.85\linewidth]{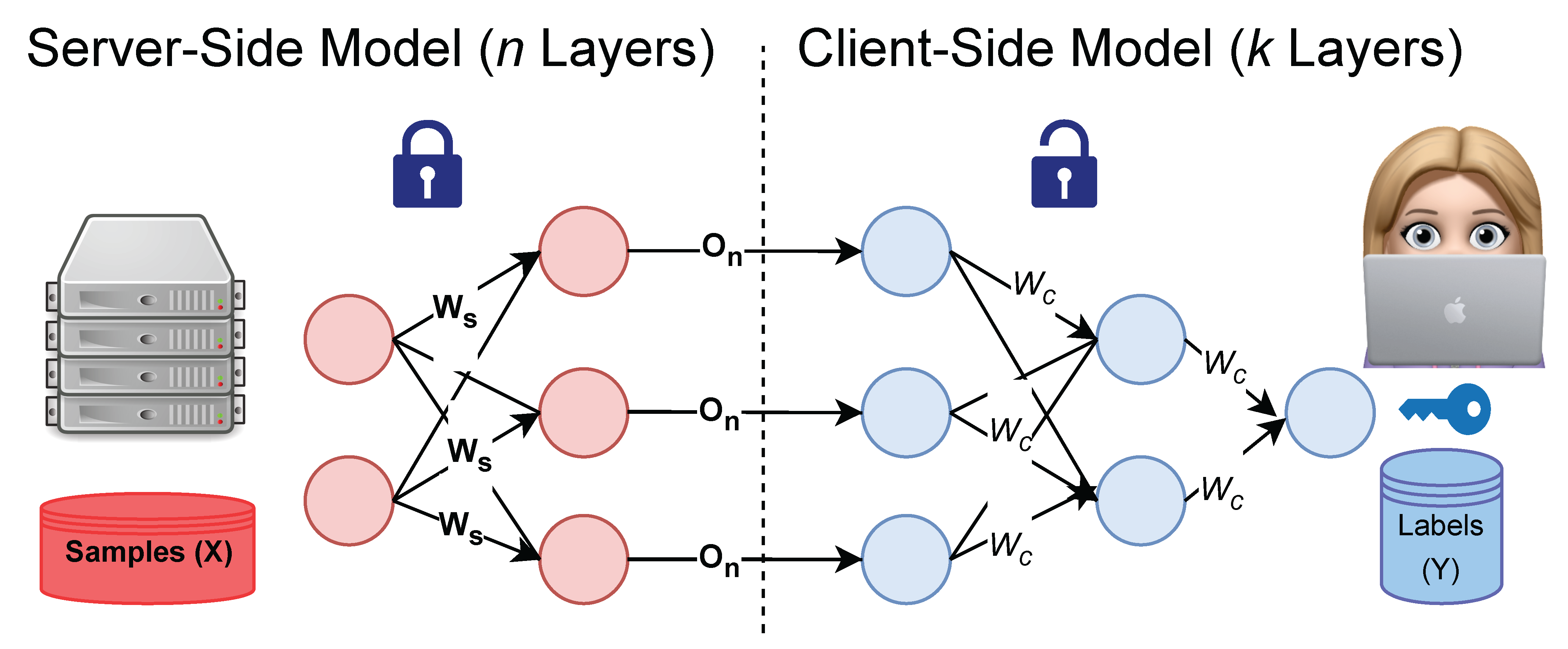}
        \label{fig:cure-model}
    }
    \caption{\revised{(a) Vanilla SL and (b) \sys's inverted SL model.}}
    \label{fig:vsl-vs-cure}
\end{figure}

\noindent\textbf{{Threat Model.}}
We consider a semi-honest \revised{threat model, which remains the most practical and widely accepted assumption in HE-based SL frameworks, including~\cite{khan2023split, nguyen2023split, khan2023love}.} \revised{Most SL systems also} operate in a single-client, single-server setting~\cite{shabbir2025taxonomy}, \revised{where both parties follow the protocol honestly without collusion. However, the server may passively attempt to infer sensitive information from exchanged messages or model parameters. This honest-but-curious model is common in collaborative ML and enables the design of more efficient cryptographic protocols~\cite{Sagar2021Confidential}.} \revised{Our threat model also reflects realistic deployment scenarios. In particular, the semi-honest model achieves an effective trade-off between security and efficiency. It provides strong confidentiality guarantees via semantically secure HE schemes such as CKKS, while avoiding the substantial computational and communication overhead incurred by malicious-security mechanisms (e.g., zero-knowledge proofs or verifiable computation). We further discuss potential extensions of \sys to the malicious setting in Section~\ref{sec:security}.}

\descr{Motivation.} \revised{In \sys, the client owns the data but outsources storage and computation to the server, which holds only encrypted model parameters and operates solely on ciphertexts. This design ensures that the server learns nothing about sensitive label information, crucial in clinical and proprietary settings, while maintaining practical performance. Because all server-side computations are performed on encrypted data, \sys removes the main attack surfaces targeted by input-reconstruction and membership-inference attacks~\cite{shokri2017membership,zhu2019deep,pasquini_unleashing_2021,melis2019exploiting}, as well as any other attack that rely on leaked gradients, weights, input, or output, preventing the server from extracting any meaningful information.} We present a formal security definition and proof in Section~\ref{sec:security}.

\revised{
In the inverted SL setting, the server may store or process the feature data $X$, either in \textit{encrypted} or \textit{plaintext} form, while the labels $Y$ represent the most sensitive information kept by the client. To protect this information, \sys encrypts \textbf{all server parameters} throughout learning  so that the server gains no knowledge about the labels or their influence on model updates. This design also reduces the client’s computational load, as it needs to store only the labels and the final few layers of the model.

Thus, \sys is particularly suitable for settings where clients outsource data storage and portions of training, such as large-scale genomic datasets where the size of raw data (storage costs) can be dominating. Clients may offload data to the server while keeping labels private.  In some cases, transferring such labels may even be restricted by regulation. An example is Autism Spectrum Disorder (ASD)~\cite{de2014synaptic,satterstrom2020large,norman2019st}: Although genomic features (e.g., ASD-related data) may be stored unencrypted, they generally do not disclose whether an individual is affected. The labels, however, directly encode this sensitive information. Because these labels are crucial for studying complex traits shaped by many genomic variants, they require strong protection and \sys enables this by design. We note that \sys also provides optional sample privacy by encrypting client data before sending it to the server. This offers a tunable trade-off between security and computational cost based on application needs. Finally, our motivation for the resource-constrained client setting is supported in Section~\ref{sec:model_time_latency}, where we compare \sys against fully client-side plaintext training.}

\subsection{Overview of \sys}\label{System Overview}
We propose a novel framework, \sys, designed to enable collaborative machine learning across client and server with asymmetric computational resources. We employ HE, in particular the CKKS scheme (see Section~\ref{sec:CKKS}), to allow computations to be performed on encrypted data, ensuring that sensitive information remains secret and eliminating attacks via communicated values throughout the training process. \revised{We present the overall workflow of \sys in comparison to vanilla SL in Figure~\ref{fig:vsl-vs-cure}. Figure~\ref{fig:vsl-vs-cure}(a) illustrates the traditional vanilla SL setup, while Figure~\ref{fig:vsl-vs-cure}(b) depicts our inverted SL design.} Throughout the paper, we denote server-side and client-side parameters with a subscript of `s' and `c', respectively. 

The server stores the data samples $(\mathbf{X})$ and performs the forward pass ($f_s(\cdot)$) up to $n$ layers with server-side model parameters $\mathbf{W}_s$. The encrypted output $(\mathbf{O_n})$ is sent to the client, who then decrypts it and completes the forward pass ($f_c(\cdot)$) of the remaining $k$ layers, denoted as $W_c$. Note that $\mathbf{W}_s$ denotes the server-side weight matrix while $\mathbf{w_i}$ are the column/row entries (vectors) of the matrix. 
The client, which holds the labels (\( Y \)), computes the loss $(J)$ and its gradients ($g_{W_s}$ and $g_{W_c}$), updating client weights $W_c$ locally and sending the encrypted $(\mathbf{g_{W}}_s)$ to the server. The server updates its parameters in an encrypted fashion. This process ensures that (optionally the data \( X \)  and) the labels \( Y \) remain confidential, adhering to the objectives of our inverted SL framework.


Our protocols' security relies on the premise that the server, despite observing encrypted gradients communicated during the training, cannot deduce the underlying labels better than random guessing, provided that the HE scheme effectively makes the encrypted values indistinguishable from random.
\sys is designed to ensure that all interactions and computations are conducted securely to maintain data and/or label privacy throughout the process. This approach protects sensitive information while enabling scalable, efficient distributed learning when participants differ in computational power or data sensitivity.

\begin{algorithm}[t]
\caption{CURE Training Phase}
\label{alg:CURE}
\begin{algorithmic}[1]
\Statex Server has (encrypted) data $\mathbf{X}_{[1, 2, \dots, m]}$, encrypted initial weight matrix $\mathbf{W}_s$, and the client's public key $\textsf{PK}_c$.
\Statex {Client has labels $Y$.}
    \For{epoch $=1 \rightarrow e$}
        \For{$\mathbf{X}$ $\in$ $\mathbf{X}_{[1, 2, \dots, m]}$}
            \State \textbf{Server performs encrypted  forward pass:}
            \State \quad $\mathbf{O}_n \gets \textsf{Eval}_{\textsf{PK}_c}^{f_s}(\mathbf{W}_s, \mathbf{X})$
            \State \quad $\text{Send } \mathbf{O}_n$ \text{to client} 
            \State \textbf{Client works on plaintext:}
            \State \quad $O_n \gets \text{Dec}_{\textsf{SK}_c}(\mathbf{O}_n)$
            \State \quad  $\hat{Y} \gets f_c(W_c,O_n)$
            \State \quad $J \gets \text{Loss}(\hat{Y}, Y)$
            \State \quad Compute gradients $g_{W_s}, g_{W_c}$
            \State \quad $W_c \gets \text{Update}_c(W_c, g_{W_c})$
            \State \quad $\mathbf{g_{W}}_s \gets \text{Enc}_{\textsf{PK}_c}(g_{W_s})$
            \State \quad Send $\mathbf{g_{W}}_s$ to server
            
            \State \textbf{Server performs encrypted backpropagation:}
            \State \quad $\mathbf{W}_s \gets \textsf{Eval}_{\textsf{PK}_c}^{\text{Update}_s}(\mathbf{W}_s,\mathbf{g_{W}}_s)$
        \EndFor
    \EndFor
\end{algorithmic}
\end{algorithm}
\subsection{\sys's Design:}
\subsubsection{{Initialization}} This phase of \sys, as detailed in Algorithm~\ref{Innitialization} in Appendix~\ref{sec:supplementaryAlgo}, is crucial for setting up the necessary cryptographic keys and model parameters for secure and efficient training. Initialization begins by defining the split model architecture, where \( L_s \) represents the server-side layers \([l_1, l_2, \ldots, l_n]\) and \( L_c \) represents the client-side layers \([l_{n+1}, \ldots, l_{n+k}]\). Thus, $l_n$ denotes the split layer index; for example, $l_2$ indicates that the network is split after the second layer between client and server. 
The client and the server randomly initialize their weights through the $GenRandomWeights$ function that randomly initializes the weight matrices for a set of layers (Lines 4 and 9). The client also generates a pair of public and secret keys \((\textsf{PK}_c, \textsf{SK}_c)\) using \text{KeyGen} operation of HE scheme (Line 5), and then sends the public key \((\textsf{PK}_c)\) to the server (Line 6). The server encrypts its weights (\( \mathbf{W}_s \)) using \(\textsf{PK}_c\) (Line 10). Thus, initialization ensures that the server-side weights are encrypted before any data exchange, maintaining privacy from the outset.
If the server does not know the input data $X$, which is only known by the client, then the client encrypts it locally to prepare for training, and sends the ciphertext $\mathbf{X}$ to the server.

\subsubsection{{Training}} \sys's training algorithm is detailed in Algorithm~\ref{alg:CURE}. 
First, the server performs the forward pass of $n$ layers under encryption, either on encrypted data $\mathbf{X}$, or on plaintext data $X$, depending on the application. We note that in the latter case, \sys only protects label confidentiality. At this step, the server computes a forward pass $f_s(\cdot)$ on its model portion using the encrypted weights $\mathbf{W}_s$ and the batch $\mathbf{X}$ (Line 4), producing an encrypted output $\mathbf{O}_n$ of $n$ layers, which is then sent to the client (Line 5). The client then decrypts $\mathbf{O}_n$ using its secret key $\textsf{SK}_c$ (Line 7), and performs a forward pass $f_c(\cdot)$ on its $k$-layer model portion using $O_n$ and local weights $W_c$ (Line 8), resulting in the predicted output $\hat{Y}$. The loss $J$ is computed using $\hat{Y}$ and the true labels $Y$ (Line 9). Gradients for the client side $(g_{W_c})$ and server side $(g_{W_s})$ are computed (Line 10). \revisedTwo{Note that Here, $g_{W_s}$ denotes the gradient $\frac{\partial J}{\partial O_n}$ at the split index, which the server uses alongside its stored encrypted activations to compute its actual weight updates under HE.} The client updates its weights $W_c$ using its gradient $g_{W_c}$ (Line 11), encrypts the server gradient $g_{W_s}$ with $\textsf{PK}_c$ (Line 12) as $\mathbf{g_{W}}_s$, and sends the result to the server (Line 13). Finally, upon receiving $\mathbf{g_{W}}_s$, the server updates its $\mathbf{W}_s$ accordingly (Line 15). This process repeats for each batch and continues for the predefined number of epochs $(e)$, ensuring efficient and secure training of the model through collaborative computation between the client and server. The client computation performed on plaintext (Lines 7 to 11) is referred to as $UpdPlain$ for convenience.

\subsection{Homomorphic Operations}
This section outlines our approach to homomorphic operations and their corresponding implementation details.
\subsubsection{Homomorphic vs.\ Plain Evaluation Modes}%
\label{sec:he_vs_plain}
In our framework, each neural network layer supports two modes of operation: a homomorphic mode for encrypted computations and a plaintext mode for standard computations. This dual-mode design guarantees functional equivalence of the underlying mathematical operations—up to the noise inherent in the CKKS scheme—across both settings. Such equivalence is essential for verifying the correctness of the cryptographic protocols and enables flexible placement of the server–client split point at any layer of the network.
 

\subsubsection{Packing}%
\label{sec:packing}

In CKKS, each ciphertext carries up to $N/2$ real slots, which we exploit to co-locate data and amortize HE costs. In \sys, we support SIMD packing in various parts of the implementation to achieve operability using homomorphic operations:

\descr{1. Data Packing for Linear Layers}
To perform matrix-vector multiplication $y = Wx + b$ homomorphically, we employ a row-wise packing strategy. During a setup phase, each row of the weight matrix $W$ is encoded into a dedicated ciphertext. The elements of the $j$-th row, $W_{j,*}$, occupy the initial slots of this ciphertext. This layout is designed for maximizing computational throughput 
(see Appendix~\ref{sec:supplementaryPacking}). This layout is designed to compute the dot product $W_{j,*} \cdot x$ homomorphically using a single ciphertext-ciphertext multiplication followed by a tree-like summation of the resulting slots.

For the backward pass, which requires multiplication by the transposed weight matrix $W^T$, weights are also pre-encoded in a column-wise fashion into separate plaintexts. Additionally, a set of one-hot plaintext vectors is prepared to facilitate the extraction and placement of specific slots during the computation.

\descr{2. Channel-Block Packing for Convolutional Layers}
For convolutional layers, we introduce a channel-packing technique to maximize the SIMD throughput. The goal is to compute multiple output channels in parallel within a single ciphertext. Given an input with $C_{in}$ channels and a kernel of size $\kappa \times \kappa$, the number of slots required to process a single receptive field for one output channel is $D_{patch} = C_{in} \cdot \kappa^2$. We can therefore pack $G_{size}$ output channels into a single ciphertext, where: $
G_{size} = \left\lfloor \frac{N/2}{D_{patch}} \right\rfloor$. 
During a pre-computation phase, the kernel weights are encoded into a series of plaintext masks. For each spatial position $(dy, dx)$ in the kernel, a specific mask is constructed. This mask contains the weights for that position across all input channels, structured in a block-diagonal format where each block corresponds to a group of $G_{size}$ output channels. This layout lets one homomorphic multiplication apply the kernel weights for a given spatial offset to all corresponding inputs across a whole group of output channels at once. This significantly reduces the number of required multiplications and rotations during the forward pass. A similar mask set is used to efficiently compute gradients in the backward pass.

By tailoring the number of slots to the layer’s fan-in (columns) or group of filters (channels), we minimize the number of ciphertexts, reduce rotations, and maximize SIMD throughput. See Appendix~\ref{sec:supplementaryPacking} for examples of packing implementations. Next, we explain the homomorphic evaluation of these linear and convolutional layers. 

\subsubsection{Homomorphic Evaluation of Linear Layers}%
\label{sec:he_linear}

Our linear layer provides a unified API for forward, backward, and update operations in the encrypted setting. Below, we detail the forward pass, backward pass, and weight update operations. 

\descr{Forward Pass.}
We compute the operation $y = Wx+b$ and produce single ciphertext where the $j$-th slot holds the $j$-th element of the output vector $y$.  A concrete breakdown of our one-level and n-level linear layer implementations, which realize this process, is provided in Appendix~\ref{sec:supplementaryOneLevel} and Appendix~\ref{sec:supplementaryNLevel}, respectively. We summarize the sequence of operations for each output neuron $j$:
\begin{enumerate}[leftmargin=*]
    \item \textbf{Homomorphic Dot Product:} The input ciphertext, containing the vector $x$, is multiplied element-wise with the pre-encoded ciphertext for the $j$-th weight row. The resulting ciphertext is then summed internally using a tree-based pattern of optimized rotations and additions. This collapses the element-wise products into a single slot (slot 0), yielding the encrypted dot product $W_{j,*} \cdot x$.
    \item \textbf{Result Placement:} The encrypted dot product, located in slot 0, is moved to the $j$-th slot via a rotation. A pre-computed one-hot plaintext mask is then applied to zero out other slots.
    \item \textbf{Output Assembly:} The resulting ciphertexts---each containing a single dot product at the correct output position---are added together. This assembles a final output ciphertext where the first $|l_{i+1}|$ slots contain the parallel dot-product results.
    \item \textbf{Bias Addition:} The plaintext-encoded bias vector is added to the assembled ciphertext to produce the final result.
\end{enumerate}

We support two packing strategies for the dot‐product:
\begin{itemize}[leftmargin=*]
  \item \textbf{Batch multiplication ($\odot$):} Pack multiple weight columns into each mask, perform one \texttt{MulNew} and tree‐sum per column‐group.
  \item \textbf{Scalar multiplication ($\otimes$):} Mask and sum one column at a time.
\end{itemize}
The ratio $\frac{N/2}{|l_2|}$ decides the preferred method (see Section~\ref{sec:estimator}). Both appear in one-level batch multiplication and one-level scalar multiplication methods (Appendix~\ref{sec:supplementaryOneLevel}).

\descr{Backward Pass.}
The backward pass homomorphically computes the weight ($g_{W_s}$) and the input gradients ($g_X$). Note that all gradients, including $g_X$, are denoted as $g_{W_s}$ in Algorithm 1 for clarity.
\begin{enumerate}[leftmargin=*]
    \item \textbf{Weight Gradients ($g_{W_s}$):} The gradient for the $j$-th weight row is the outer product of the $j$-th element of the output gradient and the cached input vector from the forward pass. This is computed by first isolating the $j$-th gradient element from the incoming gradient ciphertext. This element is then homomorphically broadcast across a new ciphertext, which is then multiplied by the cached input ciphertext. This process is repeated for each row of the weight matrix.
    \item \textbf{Input Gradients ($g_X$):} The input gradient is computed by multiplying the transposed weight matrix $W^T$ by the output gradient ciphertext. This is performed by multiplication with pre-encoded plaintext columns of the weight matrix. The results are then aggregated using a tree-sum to produce the final input gradient ciphertext.
\end{enumerate}

\descr{Update.}
The server updates its encrypted weights using the computed gradients. Each encrypted weight gradient ciphertext is multiplied by a plaintext encoding of the learning rate $\alpha$. The resulting ciphertext is then subtracted from the corresponding encrypted weight ciphertext, performing a secure stochastic gradient descent.

\subsubsection{Homomorphic Convolution}%
\label{sec:conv}
Our homomorphic 2D convolution maps the sliding-window operation to ciphertext manipulations by packing each input channel into a row-major ciphertext. 

\descr{Forward Pass.} The convolution with a $\kappa \times \kappa$ kernel is performed as follows:
\begin{enumerate}[leftmargin=*]
    \item \textbf{Input Alignment:} For each position $(dy, dx)$ in the kernel, the entire input ciphertext is rotated. The rotation amount is determined by the input feature map's width ($W_{in}$). The total rotation is calculated as $-(dy \cdot W_{in} + dx)$ slots. This rotation aligns the input data corresponding to that kernel position across all receptive fields simultaneously.
    \item \textbf{Masked Multiplication:} The rotated input is multiplied by a pre-encoded plaintext mask. This mask contains all the kernel weights for the given position $(dy, dx)$, packed in a block-diagonal structure that aligns with groups of output channels. This single multiplication applies the kernel weights for one spatial position to the corresponding input values for all output channels in parallel.
    \item \textbf{Accumulation:} The results from each masked multiplication are accumulated via homomorphic addition. This step effectively sums the products over the entire receptive field, completing the dot-product computation.
    \item \textbf{Bias Addition:} After iterating through all kernel positions and input channels, the plaintext-encoded bias is added to the accumulated result.
\end{enumerate}
This process yields a set of output ciphertexts, where each contains the feature maps for a group of output channels.

\descr{Backward Pass.}
The backward pass computes the gradients for the weights ($g_W$) and the input ($g_X$) gradients homomorphically.
\begin{enumerate}[leftmargin=*]
    \item \textbf{Weight Gradients ($g_W$):} The gradient $g_W$ is computed by performing a cross-correlation between the input feature maps (cached from the forward pass) and the loss gradient with respect to the convolution output. This involves multiplying the output gradient ciphertext with appropriately rotated versions of the input ciphertexts. Pre-computed gradient masks are used to efficiently collect these products according to the kernel structure and channel grouping. The results are aggregated and bias gradients are computed separately.
    \item \textbf{Input Gradients ($g_X$):} $g_X$ are computed using a process similar to transposed convolution. This involves multiplying the output gradient ciphertext with rotated versions of the kernel weights. These operations are aligned and accumulated to reconstruct the gradients for the input feature maps.
\end{enumerate}

\descr{Update.} We follow the same procedure as the homomorphic update of the linear layer.

\subsubsection{Homomorphic Average Pooling}%
\label{sec:pool}
We implement encrypted average pooling over a $p \times p$ window using a two-phase gather-and-place approach that consumes a single multiplicative level. For each non-overlapping window in the feature map:
\begin{enumerate}[leftmargin=*]
    \item \textbf{Gather Phase:} To compute the sum of all elements in a window, each element is first isolated using a one-hot plaintext mask and then rotated to a common accumulation slot (slot 0). The results of these are homomorphically added, producing a ciphertext with the window's sum in slot 0.
    \item \textbf{Place Phase:} The ciphertext containing the sum is multiplied by a plaintext encoding the scaling factor $1/p^2$ to compute the average. The result, located in 0th slot, is rotated to its final position in the output feature map.
\end{enumerate}
This procedure is repeated in parallel for all windows, producing a single output ciphertext for each input feature map.

Pooling has no trainable parameters; during backpropagation, gradients are simply scattered to contributing inputs, which in HE occurs over ciphertexts. First, each output gradient is isolated from its ciphertext using a one-hot plaintext mask. This isolated gradient is then scaled by a plaintext encoding the factor $1/p^2$, where $p$ is the pool size. Finally, this scaled value is distributed to the $p \times p$ original input slots through rotations, with each rotation moving the value to its correct input position. The final input gradient is the homomorphic sum of all these distributed values.

\revisedTwo{Non-linear activations cannot be applied directly under encryption (only polynomial functions are allowed). To address this, we employ polynomial approximations for both activation functions and their derivatives (see Appendix~\ref{sec:ActivationFunction} for details). This approach allows us to evaluate widely adopted non-linear activation functions (e.g., ReLU) while keeping the multiplicative depth to a minimum. Since these element-wise operations do not require expensive ciphertext rotations, their computational cost is negligible compared to the linear layers.} 

\subsubsection{Bootstrapping} \label{bootstrapping}
For an initial level of $L$, CKKS allows for at most $L$ multiplications to be carried out. As encrypted data undergoes multiple operations, noise accumulates, potentially making ciphertexts undecipherable. Thus, after $L$ multiplications, \(\textsf{Bootstrap}(c)\) function (see Section~\ref{sec:CKKS}) must be executed to refresh the ciphertext level, enabling continued operations. In \sys, we rely on bootstrapping operations when the combined number of encrypted layers $n$ and the degree of the activation function $d$ consumes all available levels, i.e., when $(log_2(d+1))n+n>L$. In practice, different server layers may use different approximation levels (and even different activation functions). The total number of bootstraps required during training, denoted as $\gamma$, can be expressed as:
\begin{equation}
\gamma = \frac{\sum_{i=1}^{l} \left( \frac{l_i \times l_{i+1}}{\frac{N}{2}} \right) \times (\mu ) \times (1 + d)}{L}
\label{bootstrap equation}
\end{equation}
Here, $N/2$ denotes the number of slots in the RLWE vector, and $\mu$ is the multiplicative depth. Bootstrapping is performed when the accumulated operation depth across server-side layers exceeds the multiplicative depth limit. This allows encrypted server-side computations to proceed correctly, preserving efficiency and security.
\begin{table*}[h!]
\scriptsize
\centering
\begin{tabular}{@{}l>{\raggedright}p{5cm}c>{\centering\arraybackslash}p{3cm}@{}}
\toprule
\textbf{Operation} & \textbf{Time Complexity} & \textbf{Used memory float32} & \textbf{\# Levels Used} \\ \midrule
\textbf{One-Level Scalar Multiplication} & $\sigma_{\otimes} \cdot |l_0|$ & $|c| \cdot \lceil\frac{|l_1|}{N/2}\rceil$ & 1 \\ \midrule 

\textbf{One-Level Batch Multiplication} & $\sigma_{\odot} \cdot \frac{|l_0|}{N/2}$ & $|c| \cdot \lceil\frac{2 \cdot |l_1|}{N/2}\rceil$ & 1 \\ \midrule 

\textbf{Approximated Activation Polynomial} & $\sigma_{\odot} \cdot \lceil\text{log(d)}\rceil$ & - & $\lceil\text{log(d)}\rceil$ \\ \midrule

\textbf{Server-Client Communication} & - & $\lceil\frac{|l_n|}{N/2}\rceil \cdot |c|$ & 0 \\ \midrule

\textbf{Bootstrapping} & $T_{\text{bootstrap}} \cdot \left [\sum_{l=1}^{L} \frac{|l_i| \times |l_{i+1}|}{N/2} \times \mu \times (1 + d) \right] $ & $\frac{|l_i| \times |l_{i+1}|}{N/2}$ & 0 \\ \midrule

\textbf{HE Forward Pass} & $T_{r} \cdot \left[\sum_{i=0}^{n} \left( \frac{|\bar{l}_i| \times |\bar{l}_{i+1}|}{N/2} \right) \times \log(|\bar{l}_{i+1}|)\right]$  & $ |c| \cdot  \sum_{i=0}^{n} \left( \frac{|\bar{l}_i| \times |\bar{l}_{i+1}|}{N/2} \right)$ & $n-1+ \sum_{i=1}^{n-1} \text{log($\bar{d}$)}$ \\ \midrule


\textbf{HE Back-propagation} & $T_{r} \cdot \left[\sum_{i=0}^{n} \left( \frac{|\bar{l}_i| \times |\bar{l}_{i+1}|}{N/2} \right) \times \log(|\bar{l}_{i+1}|)\right]$  & $ |c| \cdot \sum_{i=0}^{n} \left( \frac{|\bar{l}_i| \times |\bar{l}_{i+1}|}{N/2} \right)$ & $n-1+ \sum_{i=1}^{n-1} \left(\text{log$(\bar{d})$} - 1\right)$ \\ \midrule 



\textbf{HE Convolution} &
$\,C \!\left[\kappa^{2}(T_r + \sigma_{\odot}) + \sigma_{\otimes}\right]$ &
$\lceil C \rceil \!\cdot\! |c|$ &
$n-1+ \sum_{i=1}^{n-1} \text{log($\bar{d}$)}$  \\ \midrule
\textbf{HE Average Pooling} &
$\bigl(p^{2}\!+\!1\bigr)T_r + p^{2}\sigma_{\otimes}$ &
$|c|$ &
$1$ \\ 

\bottomrule 
\end{tabular}

\caption{Complexity analysis of CURE's fundamental operations for SL training with a cyclotomic ring size $N$. $\sigma_{\otimes}$ and $\sigma_{\odot}$ denote the execution time of scalar HE multiplications and HE batch multiplications. $T_{r}$ and $T_{\text{bootstrap}}$ indicate the execution time of a rotation and bootstrapping, respectively. $|c|$ denotes the length of an RLWE vector ciphertext, $\bar{|l_i|}$ the smallest power of two exceeding the size of the $i^{\text{th}}$ layer, $d$ the polynomial degree of the approximated activation function, and $\text{dot}(\cdot,\cdot)$ the time for a dot product of given matrix dimensions. }

\label{tab:complexity_table}
\vspace{-2em}
\end{table*}

\subsection{Estimator for Server-Client Optimization}\label{sec:estimator}

In this section, we develop a novel estimator to enhance \sys's utilization. Building on its theoretical foundation, we implement an advisor function to optimize the server-client split in \sys. This function aims to determine the optimal partitioning of neural network layers between server and client while optimizing HE operations, training latency, memory usage, and bootstrapping frequency. The advisor takes key properties such as the desired training time ($T_d$), computer specifications to calculate the latency of rotations, the depth of the multiplicative circuit ($\mu$), the depth of the additive circuit ($\rho$), the total number of bootstrapping operations ($\gamma$) calculated in Formula~\ref{bootstrap equation}, and the network bandwidth ($\mathcal{O}_c$) available between the client and server. We focus on these parameters as our preliminary experiments suggest they are the most dominant factors. Rotations significantly affect training latency, making them crucial for time efficiency, while $\mu$ and $\rho$ influence accuracy and the need for bootstrapping. Consequently, we provide \sys-advisor that can provide recommendations for scenarios where the client's computational power is limited, communication bandwidth is constrained, or the precision impact from the server side is critical. We present the runtime analysis, comparing the estimates from the advisor function with the actual runtime observed in Section~\ref{sec:estimatorEval}. Finally, we introduce \sys's complexity analysis in Table~\ref{tab:complexity_table} to facilitate the estimator function and its implementation. 

\subsubsection{Server-Side Computational Complexity}

The server-side forward pass involves homomorphic matrix-matrix multiplication and one-level operations, introducing computational overhead due to HE. The computational complexity is largely dominated by rotations and HE multiplications. The total number of encrypted matrix-matrix multiplications for one forward pass is given by:
\begin{equation}
\sum_{i=0}^{n} \left( \frac{|\bar{l}_i| \times |\bar{l}_{i+1}|}{N/2} \right) \times \log(|\bar{l}_{i+1}|)
\label{eq:linear_layer_estimator}
\end{equation}
where $N/2$ is the number of slots, and $|\bar{l}_i|$ is the smallest power of two greater than the size of the $i^{th}$ layer. This equation defines the time complexity of the server's forward pass, taking into account the logarithmic cost of rotations.

For other layers, such as convolutional and pooling layers, the complexity analysis similarly considers the specific HE operations required, as detailed in Table~\ref{tab:complexity_table}. This includes the cost of rotations for input alignment, masked multiplications for weight application, and accumulations. The server-side complexity estimation takes into account the number of samples processed, batch size, and the packing strategies defined for each layer type, similar to the principle illustrated in the linear layer example.

Additionally, the server also tracks the cumulative homomorphic depth up to the current layer. This depth is crucial because when $\delta_{\text{current}}$ exceeds the maximum permissible depth $\delta_{\text{max}}$, bootstrapping is triggered, incurring additional cost. The cumulative depth $\delta_{\text{current}}^{(i)}$ for each server layer $i$ is calculated as:
\begin{equation}
\delta_{\text{current}}^{(i)} = \delta_{\text{current}}^{(i-1)} + \left(\frac{l_i \times l_{i+1}}{\frac{N}{2}}\right) \times (\mu + \rho) \times (1 + d)
\end{equation}
Here, $d$ is the degree of the activation function and $\mu$ and $\rho$ are the depths of the multiplicative and additive circuits, respectively. The server performs $\delta_{\text{current}}$ calculation to determine when bootstrapping is necessary. For simplicity in notation, we assume all activation functions have a degree of $d$. Selecting the maximum degree as $d$ among the activation functions, our analysis establishes an upper bound on the overall complexity. The server estimator ultimately provides its estimation as $T_{server}$, accounting for both the forward and backward pass, as well as bootstrapping if needed.
\begin{equation}
\scriptsize
T_{\text{server}} \approx 2 |X| \cdot \left[ \sum_{i=0}^{n} (\text{Cost}_{\text{layer}_i}) + \text{Cost}_{\text{bootstrapping}} \right]
\label{eq:server_return}
\end{equation}
where $\text{Cost}_{\text{layer}_i}$ encompasses the cost of HE rotations for layer $i$, (e.g., $\frac{|\bar{l}_i| \times |\bar{l}_{i+1}|}{N/2} \times \log(|\bar{l}_{i+1}|)$ for linear layers) and $\text{Cost}_{\text{bootstrapping}}$ accounts for the cost of refreshing ciphertexts when the multiplicative depth is exceeded.

\subsubsection{Client-Side Complexity and Memory Constraints}

On the client side, we aim to minimize computational overhead while meeting memory and time constraints. The complexity of these operations is primarily driven by the dot products required for matrix multiplications, which can be calculated as:
\vspace{0.2em}
\begin{equation}
\sum_{i=n}^{m} \text{dot}\left(|l_i| \times |l_{i+1}|, |l_{i+1}| \times |l_{i+2}|\right)
\end{equation}
\vspace{0.2em}
where $\text{dot}(\cdot,\cdot)$ represents the dot product of matrices on the client side with the specified sizes, and $|l_i|$ refers to the size of the $i^{th}$ layer. It is important to note that operations are performed in plaintext on the client side. Additionally, the client must ensure that the memory required for storing encrypted weights does not exceed available capacity. The memory required for storing ciphertexts is proportional to $|c| \cdot \left\lceil\frac{|l_n|}{N/2}\right\rceil$ where $|c|$ denotes the size of one RLWE vector under the given parameter set, and $|l_n|$ represents the size of the last layer processed by the client. This expression represents the theoretical upper bound for memory consumption on the client side. Together, the above calculations enable us to quantify the primary client-side constraints: computational load and memory usage. Client estimator provides its estimation as $T_{\text{client}}$.
\begin{equation}
T_{client} = 2|X| \times \sum_{i=n}^{m} \text{dot}\left(|l_i| \times |l_{i+1}|, |l_{i+1}| \times |l_{i+2}|\right)
\end{equation}
\subsubsection{Data Transfer and Communication Overhead}

Another key factor in the advisor function is the communication overhead, determined by the number of ciphertexts transferred. Per epoch, the server sends:
\begin{equation}
T_{\text{comm}} = \frac{|X| \times |l_n|}{N/2} \times |c|
\label{eq:memory_calculation}
\end{equation}
where $|X|$ is the number of data samples, $|l_n|$ is the size of the last layer processed by the server, and $|c|$ is the memory size of one ciphertext. By minimizing $|l_n|$, our advisor function can effectively reduce the number of ciphertexts transferred, thereby optimizing communication and enhancing overall efficiency.

\subsubsection{Optimal Partitioning Strategy}

The advisor function combines computational, memory, and cryptographic constraints from both server and client to determine the optimal split. It computes the server’s maximum feasible split index based on rotations and memory limits, then compares it with the client’s minimum feasible index to balance efficiency and memory use.

For each feasible split index where the server and client estimators agree, the advisor calculates the total computational time (including rotations and bootstrapping) and communication overhead. It then selects the partition that minimizes training time while satisfying the user’s memory and time constraints. The advisor's final recommendation is guided by the following objective: $
\text{Minimize} \left(T_{\text{server}} + T_{\text{client}} + T_{\text{comm}} \right)$ 
where $T_{\text{server}}$, $T_{\text{client}}$, and $T_{\text{comm}}$ represent the respective training times for the server, client, and communication phases. By accounting for the complexity of each operation, as outlined in Table~\ref{tab:complexity_table}, the advisor function offers a theoretically sound method for optimizing the partitioning of NN layers in \sys. It also reports which constraints --server time, client time, or communication-- exceed their thresholds.

\subsection{Security Definition and Proof}\label{sec:security}
We start by observing that previous solutions combining SL with HE \cite{Pereteanu_Alansary_Passerat_Palmbach_2022,khan2023love,khan2023split,khan2023more} do not provide any formal security definition and proof. In contrast, we state our guarantee formally and sketch a proof. Our definition of the ideal functionality $\mathcal{F}_{InvSL}$ in Functionality 1 is inspired by~\cite{mohassel2017secureml}, but adapted to the inverted SL setup. Observe that the functionality does not return any output to the server, which means the server learns no useful information. The client, on the other hand, obtains intermediate outputs at each epoch, but is considered as honest in our case, since (s)he owns the data and labels. Note that since we consider the semi-honest case, we did not complicate the functionality with aborts.

\begin{figure}[!htp]
\begin{functionality}{Inverted SL Ideal Functionality $\mathcal{F}_{InvSL}$} \label{functionality:secureSL}

\textit{\textbf{Parameters:}} 
Number of epochs $e$.

\begin{enumerate}
    \item Receive the model description from the participants. In particular, receive the forward pass $f_s$ and backpropagation $\text{Update}_s$ from the server, and the plaintext forward and backward pass combination $UpdPlain$ from the client.
    \item Depending on the configuration, receive the input $X$ either from the client or the server, and receive the labels $Y$ from the client.
    \item Initialize weights $W_s$ and $W_c$ using $GenRandomWeights$.
    \item For each epoch, up to $e$ epochs, compute
    $$
    O_n^e = f_s(W_s,X)
    $$
    $$
    W_s = \text{Update}_s(UpdPlain(O_n^e, Y))
    $$
    and send $O_n^e$ to the client.
\end{enumerate}

\end{functionality}
\vspace{-1em}
\end{figure}


Functionality 1 ensures that the server learns nothing useful during training. Our security theorem and proof are below.

\begin{theorem}
Assuming that the underlying HE scheme provides CPA security, then \sys\ realizes $\mathcal{F}_{InvSL}$ against a semi-honest server.
\end{theorem}

\textit{Proof: }
Note that to prove ideal–real indistinguishability against a semi-honest adversary, it suffices to construct a simulator that reproduces indistinguishable outputs and views given the adversary’s inputs. In our case, the server's inputs include the model description in the form of the forward pass $f_s$ and backpropagation $\text{Update}_s$, as well as optionally the samples $X$. The simulator $\mathcal{S}$ sends these inputs to the ideal functionality $\mathcal{F}_{InvSL}$ and receives back nothing, since the server has no output. In our scenario of SL, the only values visible to the server are plaintext or encrypted samples ($X$ or $\mathbf{X}$), encrypted weights ($\mathbf{W}_s$), and encrypted gradients ($\mathbf{g_{W}}_s$). Initially, assume that the data $X$ is known by the server. In that case, the simulator $\mathcal{S}$ simulates the view of the adversary simply by outputting randomly encrypted values for $\mathbf{W}_s$ and $\mathbf{g_{W}}_s$. In the case where the client knows the data $X$ and only sends its encryption $\mathbf{X}$ to the server during initialization (and therefore the server does not have plaintext access to the data samples), the simulator encrypts junk instead of samples and outputs $\mathbf{X}$, in addition to the randomly encrypted junk $\mathbf{W}_s$ and $\mathbf{g_{W}}_s$, to simulate the adversary's view.

This simulator produces a view and output that is indistinguishable from those of the adversary, given that the HE scheme is CPA-secure. A full indistinguishability proof could use a hybrid argument, where ciphertexts would partly be based on real data and based on junk. Given the public key for the encryption scheme, all these ciphertexts can be generated as necessary. The challenge ciphertext received (either real or junk) would be put to a randomly picked location $j$ within the encrypted weights/gradients, and the hybrid encrypted weights/gradients would be shared with the distinguisher. Until $j-1$ all ciphertexts would be based on real values. Up from $j+1$ all ciphertexts would be based on junk (random) data. At location $j$ we will have the challenge ciphertext that we received (either real or junk). If the distinguisher acting as the server were able to distinguish these from real ciphertexts, it would break the CPA security of the underlying HE scheme. On one end, the hybrid would be all real ciphertexts, and on the other end, the hybrid would be all junk ciphertexts. Since there are only polynomially many values (gradients, weights, data samples), distinguishing each neighboring hybrid with negligible probability (due to the CPA security of the underlying HE solution) would mean distinguishing the two end-hybrids overall with negligible probability. Consequently, \sys\ ensures that no information is leaked to the server as the simulator simply encrypts random junk and realizes $\mathcal{F}_{InvSL}$.


Observe that even when the data $X$ is not encrypted on the server side, after the first forward-backward pass, an honest-but-curious server still loses any correlation between the encrypted model weights $\mathbf{W}_s$, encrypted gradients $\mathbf{g_{W}}_s$, and the original input $X$. Functionality $\mathcal{F}_{InvSL}$ guarantees that the server gains no additional information during training. For a malicious server, however, \sys needs to be combined with verifiable computation techniques~\cite{gennaro2010verifiable}, which we leave as future work. \revised{Recent advances in zero-knowledge proofs of training~\cite{cryptoeprint:2024/162} demonstrate that it is now feasible to verify, in zero knowledge, that gradient updates are correctly computed without revealing the underlying model parameters or data. Such proofs represent a promising extension for future versions of \sys, integrating zero-knowledge verification mechanisms with HE to provide security against malicious servers.}

\revised{\paragraph{Resistance to Membership Inference, Reconstruction, and Gradient Leakage Attacks.} 

Membership inference, gradient leakage, or reconstruction attacks exploit access to plaintext outputs or gradients~\cite{shabbir2025taxonomy}.  
At a high level, since the server’s entire view consists only of ciphertexts, and the CPA-secure CKKS scheme ensures that all observed values are computationally indistinguishable from random, such attacks are effectively precluded.
In CURE, backward propagation and updates on the server are also executed over encrypted gradients, meaning that gradient inversion is not feasible. 

In summary, \sys guarantees that, during training, the server is not exposed to any plaintext information that could facilitate such attacks. All server-side parameters and intermediate values are computed under CKKS encryption. If data features are also sensitive, the client also encrypts $X$ before sending it. Note that attacks targeting the inference phase (e.g., by end users or third parties) are outside the scope of this work and would require additional secure inference mechanisms. }


\section{Experimental Evaluation}
\label{evaluation}


\vspace{-0.5em}
\subsection{Experimental Setup}
\label{sec:experimental-setup}

\subsubsection{Implementation Details}
\label{implementation details}
All encrypted operators are implemented in Go with the Lattigo library v6 on the CKKS scheme. We provide further implementation details in Appendix~\ref{sec:implementationAppendix}.

\subsubsection{Setup}
\label{experimental hardware}
All timing and scalability experiments were conducted on a server with a dual Intel Xeon E5-2650 v3 CPU (providing 40 hardware cores at 2.30GHz) and 251 GB of RAM, \revisedTwo{running Ubuntu 18.04.6 LTS. All client-side experiments were conducted on a 2023 MacBook Air with the M2 chip. The communication between the client} and server for our distributed setup is implemented using the Message Passing Interface (MPI). The accuracy experiments, which utilized our noise-injection simulation, were performed on a Google Colab instance with an NVIDIA T4 GPU.

\descr{CKKS Parameters.} We tested a range of CKKS ring sizes ($N$) by varying \(\log N\), from 13 to 16. Our parameter sets were chosen to align with the recommendations from the HE Security Standard~\cite{HEStandardPaper}, with our primary set at $\log N=13$ for most of the experiments.

\subsubsection{Datasets and Model Architectures}
\label{datasets}
We use the following canonical architectures (all with degree-3 ReLU approximation). For fully-connected (FC) networks, (a--b--c--...) denotes the number of neurons in each layer from input to output, respectively, while for convolutional layers, the notation `$a \rightarrow b$` indicates a mapping from `$a$` input channels to `$b$` output channels:
\begin{itemize}[leftmargin=*]
\item \textbf{Simple MLP \cite{Nandakumar_Ratha_Pankanti_Halevi_2019}}: An FC network with layers of (64--32--16--10).
  \item \textbf{MLP \cite{Nandakumar_Ratha_Pankanti_Halevi_2019}}: A FC network with layers of (784--128--32--10).
  \item \textbf{LeNet \cite{lecun1998gradient}} A classic convolutional stack consisting of two 2D convolutional layers (first $1 \rightarrow 6$ channels with a $5 \times 5$ kernel, then $6 \rightarrow 16$ channels with a $5 \times 5$ kernel), followed by three fully-connected layers with architecture (256-120-84-10).
  
  \item \textbf{PTB-XL CNN} Khan et al.'s model~\cite{khan2023love} is a 1D convolutional network designed for ECG classification. It comprises two 1D convolutional layers (first $12 \rightarrow 16$ channels, then $16 \rightarrow 8$ channels), each followed by a pooling layer, and culminates in a final fully-connected layer mapping to 5 output classes.
 \item \textbf{Residual blocks (ResNet) \cite{resnet}} A representative block from a ResNet model, used to evaluate performance on modern, deep architectures. It consists of an initial convolutional layer ($3 \rightarrow 64$ channels with a $7 \times 7$ kernel) followed by two residual-style convolutional layers ($64 \rightarrow 64$ channels with a $3 \times 3$ kernel).  
\end{itemize}
Unless stated otherwise, we use MNIST for the MLP, Simple MLP, and LeNet experiments, and PTB-XL for the PTB-XL CNN~\cite{khan2023love} and for residual blocks.
See Appendix~\ref{sec:supplementaryDatasets} for preprocessing details.




\begin{table}[t]
\centering
\small
\begin{tabular}{lcccc}
\toprule
\textbf{Model} & Baseline (\%) & \sys $l_n$=2 (\%) & \revisedTwo{$\varepsilon_{avg}$} & \revisedTwo{$\varepsilon_{max}$} \\
\midrule
\textbf{Simple MLP} & 94.74 & 93.86  & \revisedTwo{4.0e-07} & \revisedTwo{5.8e-07} \\
\textbf{MLP} & 97.25 & 97.19  & \revisedTwo{4.0e-07} & \revisedTwo{5.8e-07} \\
\textbf{LeNet} & 98.95 & 98.76  & \revisedTwo{8.8e-08} & \revisedTwo{1.3e-07} \\
\textbf{PTB-XL CNN} & 62.83 & 62.78  & \revisedTwo{3.5e-08} & \revisedTwo{5.0e-08} \\
\bottomrule
\end{tabular}
\caption{\revisedTwo{Accuracy with split at $l_n=2$. $\varepsilon_{avg}$: per-sample average difference among plaintext and HE vectors in training. $\varepsilon_{max}$: max error among plaintext and HE vectors. These $\varepsilon$ noise values are obtained by running the training concurrently in HE and plaintext settings for $3$ epochs. Accuracy results are reported after model convergence.}}
\label{tab:model_accuracy_updated}
\vspace{-1em}
\end{table}

\begin{figure}[t]
    \centering
    \includegraphics[width=0.9\linewidth]{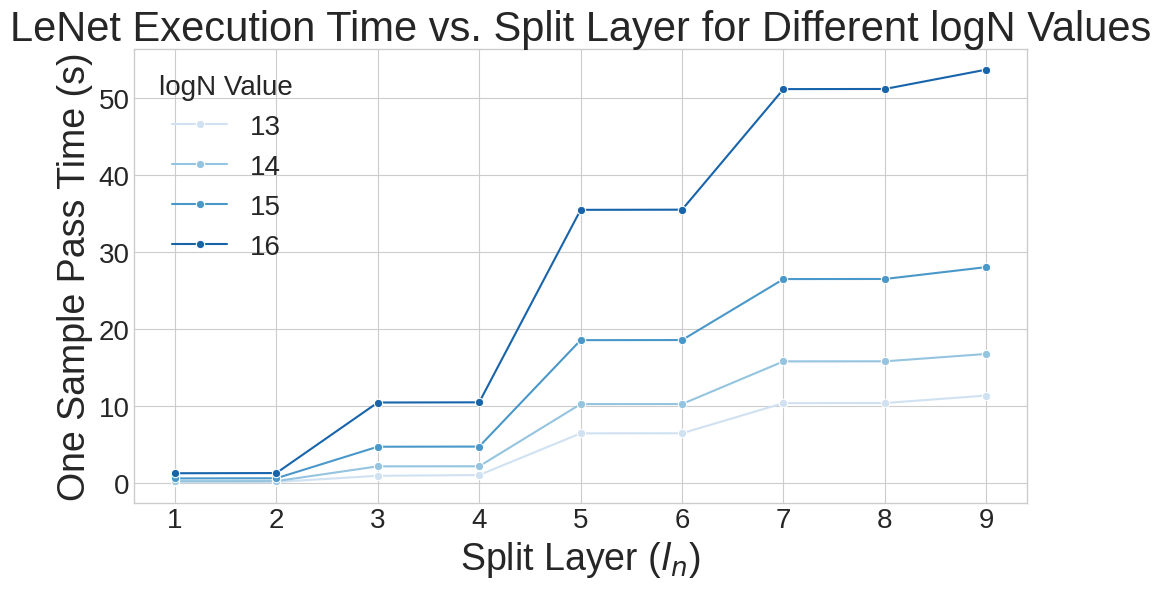}
    \caption{One sample pass time (seconds) vs. $\log N$ and split location. All benchmarks run on 40 cores.}
    \label{fig:LeNetLogNSplitLayer}
    \vspace{-1em}
\end{figure}

Finally, we detail our experimental methodolody in Appendix~\ref{sec:methodology}.

\begin{table}[t]
\centering
\small
\begin{tabular}{lccccc}
\toprule
\textbf{Model} & \textbf{$l_n$=1} & \textbf{$l_n$=3} & \textbf{$l_n$=5} & \textbf{$l_n$=7} & \textbf{$l_n$=9} \\
\midrule
\textbf{Simple  MLP} \cite{Nandakumar_Ratha_Pankanti_Halevi_2019} & 0.439 & 0.529 & 0.578 & NA & NA \\
\textbf{MLP~\cite{Nandakumar_Ratha_Pankanti_Halevi_2019}}  & 1.476 & 1.781 & 1.929 & NA & NA \\

\textbf{LeNet} \cite{lecun1998gradient} & 0.198 & 0.745 & 2.067 & 2.987 & 3.214 \\

\textbf{PTB-XL CNN} \cite{khan2023love} & 0.109 & 0.221 & 0.394 & NA & NA \\
\bottomrule
\end{tabular}
\caption{Execution times (seconds) for one sample processing with $\log N=13$ on 40 cores, across varying splits ($l_n$ denotes the last layer processed by the server). NA denotes 'Not Applicable'.}
\label{tab:40core_models_as_rows}
\vspace{-2em}
\end{table}
\subsection{Results}
\label{sec:results}

\subsubsection{Model Accuracy}
\label{sec:accuracy}
We first evaluate the impact of \sys's privacy-preserving mechanisms—namely, the degree-3 polynomial approximation of ReLU and the inherent noise from CKKS operations—on final model accuracy. \revised{We use calibrated noise simulations in our accuracy experiments. This does not bias the accuracy in favor of \sys. \revisedTwo{Crucially, rather than adding arbitrary noise, we injected the exact noise variance introduced by each HE primitive (e.g., rotation, rescaling) under our fixed parameters. Applying this empirically measured noise at each corresponding plaintext operation allowed us to accurately simulate the cumulative training error.} This simulation strategy is standard in HE-based ML research~\cite{poseidon,rhode}. To validate correctness, we compared intermediate decrypted ciphertext values against their plaintext counterparts for every layer, observing only the expected CKKS approximation error. The outcomes of simulated and fully encrypted executions were statistically indistinguishable. Therefore, the reported accuracy reliably reflects the true performance of \sys, without overstating it.}
Table~\ref{tab:model_accuracy_updated} summarizes the results from our calibrated noise-injection protocol. \revisedTwo{In addition to the accuracy metrics, we report $\varepsilon_{\text{avg}}$ and $\varepsilon_{\text{max}}$, which represent the \emph{actual measured divergence} between our HE implementation and the equivalent plaintext computation, not artificially injected noise. Specifically, $\varepsilon_{\text{avg}}$ denotes the average absolute error observed in $3$ epochs, while $\varepsilon_{\text{max}}$ shows maximum error across all samples in $3$ consecutive training epochs. These measurements, obtained by comparing CKKS ciphertext outputs directly against plaintext results, confirming that our noise-injection calibration faithfully reflects the true numerical behavior of HE under RLWE-based arithmetic.} Here, baseline refers to the plaintext implementation of our models. Our findings demonstrate that accuracy degradation remains minimal across all evaluated models. For complex architectures such as the MNIST MLP and LeNet, the accuracy drop ($\Delta$) is negligible, at -0.06\% and -0.19\%, respectively. This demonstrates that \sys can effectively preserve model utility, ensuring that the introduction of strong privacy guarantees does not significantly compromise the primary goal of the ML task.

\subsubsection{Time Latency and Scalability}
\label{sec:model_time_latency}
Here, we analyze the \sys's overall performance and its scalability with respect to \emph{the split layer index ($l_n$)}, \emph{the number of cores}, and \emph{CKKS ring size}. 

Figure~\ref{fig:LeNetLogNSplitLayer} illustrates the execution times on LeNet model across varying \textbf{split layer indices.} We use LeNet to illustrate the relationship between latency, split layers, and cryptographic parameters, as it incorporates all implemented layer types and thus captures all these interactions. We remind that $l_n$ denotes the split layer, i.e., the last layer processed by the server, from which the server sends encrypted activations to the client and receives encrypted gradients back. The results indicate that as the `split layer` position increases, latency grows substantially, reflecting the computational cost of executing more layers under HE. Table~\ref{tab:40core_models_as_rows} quantifies this trade-off across all evaluated architectures. The table confirms that for every model, deepening the split point consistently increases the computational latency. This effect is particularly pronounced for models with large initial layers, such as the MNIST MLP, where encrypting just the first layer ($l_n=1$) results in a latency of 1.476 seconds, significantly higher than the initial cost for the convolutional models. This trend peaks when the entire model is encrypted on the server (the final entry for each model) which incurs the full computational cost and highest latency.


We further examine the impact of \textbf{CKKS ring size} in Figure~\ref{fig:LeNetLogNSplitLayer}, varying $\log N$ from 13 to 16.
 Note that the ring size directly influences the security, precision, and performance of the system. The results consistently show that for any given split layer, increasing \(\log N\) leads to a notable increase in execution time. For example, for the model at $l_n = 5$, the latency increases from $6.5$\,s at $\log N = 13$ to $35{,}496$\,s at $\log N = 16$. This illustrates the computational overhead introduced by higher ring sizes and precision, underscoring this trade-off as a critical optimization parameter for practitioners.

Finally, to assess the framework's efficiency on modern hardware, we also evaluate \textbf{scalability with the number of cores}. Table~\ref{tab:logn13_cut1_allcores} reports the one-sample processing times for various models as the number of CPU cores increases from 1 to 40. For all models, the runtime decreases substantially with increasing number of cores, demonstrating near-linear speedup. For example, the LeNet model's latency drops from 7.509 seconds on a single core to just 0.198 seconds on 40 cores—a 38x speedup. This confirms that \sys efficiently exploits multi-core processors, a crucial capability for practical deployment of computationally intensive HE workloads.

\begin{table}[t]
\centering
\small
\setlength{\tabcolsep}{2.3pt}
\begin{tabular}{lrrrrrrr}
\toprule
\textbf{Model} & $c = 1$ & $c = 2$ & $c = 4$ & $c = 8$ & $c = 16$ & $c = 32$ & $c = 40$ \\
\midrule
\textbf{Simple MLP}~\cite{Nandakumar_Ratha_Pankanti_Halevi_2019}        & 16.547 & 8.379 & 4.271 & 2.187 & 1.052 & 0.551 & 0.439 \\
 \textbf{MLP}~\cite{Nandakumar_Ratha_Pankanti_Halevi_2019}   & 55.800 & 28.187 & 14.126 & 7.117 & 3.507 & 1.852 & 1.476 \\

\textbf{LeNet}~\cite{lecun1998gradient}                          & 7.509 & 3.792 & 1.901 & 0.957 & 0.471 & 0.249 & 0.198 \\
\textbf{PTB-XL CNN}~\cite{khan2023love}    & 4.130 & 2.085 & 1.045 & 0.528 & 0.271 & 0.137 & 0.109 \\
\bottomrule
\end{tabular}
\caption{One sample processing (including forward pass, backpropagation, and update) times (in seconds) obtained from benchmarks for models at $\log N=13$, $l_n = 1$, for all cores. Each core gets its own layer and key instances to reduce serial work. Each row reports runtimes using different numbers of cores ($c$), shown in the top row.}
\label{tab:logn13_cut1_allcores}
\vspace{-1em}
\end{table}

\revisedTwo{\descr{Scalability for Client.}} \revisedTwo{Finally, we examined whether using \sys offers meaningful benefits compared to the client performing full plaintext training. We note that \sys is particularly advantageous when the client lacks sufficient resources to store the dataset $X$. However, even in relatively simple settings, \sys is beneficial:}

\revisedTwo{We evaluated the end-to-end latency of three of our architectures, under two encryption parameter settings ($LogN=13$ and $LogN=14$). The results (Table \ref{tab:client_solo}) illustrate the trade-off between local computation execution and the cryptographic overhead introduced at the split index. By varying the cut layer (where the server takes over), we observe that even for a simple MLP, \sys becomes advantageous once the split is applied at layer $l_n = 3$. Similarly, for other architectures, we find that client-side computation drops favorably when the split occurs at $l_n = 2$ or $l_n = 3$. In summary, the client's total execution time is influenced by the encryption and decryption overhead at the split index. By strategically selecting the split index based on the layer's input dimension (fan-in), \sys maximizes packing density, thereby reducing the number of required ciphertexts and the associated cryptographic costs. Furthermore, even in scenarios where the computational advantage is marginal, offloading remains architecturally significant. Deferring layers to the server alleviates client-side memory constraints by removing the need to store the full model parameters and allows for secure data storage on the server side.}

\revisedTwo{\begin{table}[t]
\centering
\resizebox{\columnwidth}{!}{
\begin{tabular}{l l c c}
\toprule
\textbf{Model} & \textbf{Split Index} & \textbf{Time (ms) ($LogN$=13)}  & \textbf{Time (ms) ($LogN$=14)}  \\
\midrule
\multirow{3}{*}{\textbf{MLP}} 
 & Baseline & 5,901.84 & 5,901.84 \\
 & $l_n=$ 3 & 5,623.58 & 5,563.81 \\
 & $l_n=$ 5 & 2,350.17 & 2,335.23 \\
\midrule
\multirow{7}{*}{\textbf{LeNet}} 
 & Baseline & 23,021.99 & 23,021.99 \\
 & $l_n=$ 2 & 37,052.64 & 35,439.52 \\
 & $l_n=$ 3 & 15,415.45 & 14,937.20 \\
 & $l_n=$ 5 & 7,978.32 & 7,858.75 \\
 & $l_n=$ 7 & 4,749.25 & 4,693.20 \\
 & $l_n=$ 9 & 2,301.07 & 2,261.84 \\
\midrule
\multirow{5}{*}{\shortstack[l]{\textbf{Khan et al.}}} 
 & Baseline & 22,020.71 & 22,020.71 \\
 & $l_n=$ 2 & 46,260.19 & 43,531.27 \\
 & $l_n=$ 3 & 27,244.62 & 25,879.74 \\
 & $l_n=$ 4 & 12,966.00 & 12,286.28 \\
 & $l_n=$ 5 & 5,571.66 & 5,231.80 \\
\bottomrule
\end{tabular}
}
\caption{\revisedTwo{End-to-end latency is compared across model architectures for varying split indices and HE parameters ($\log N$). The “Baseline” corresponds to fully local, plaintext training on the client.}}
\label{tab:client_solo}
\vspace{-1em}
\end{table}}

\subsubsection{Large Models}
We evaluated our framework on a ResNet block to assess the scalability and performance trade-offs inherent in applying HE to large, modern architectures. The ResNet~\cite{resnet} block evaluated in our experiments represents a fundamental component of modern deep neural networks. Implementing this block in \sys provides a consistent baseline for fair performance comparisons and for analyzing noise propagation in residual architectures.

\descr{Accuracy.} To quantify the impact of noise in large models, our noise injection simulation measured a final standard deviation of 1.414 after a single pass on the ResNet block. This result was derived from a simulation that models an input with a normalized noise standard deviation of 1.0, adds a fixed variance for each of the four main HE operations within the block (two convolutions and two batch normalizations), and calculates the final deviation after the skip connection which is a network design element that bypasses layers to directly add an earlier layer's output to a later layer's output. Quantified noise growth per block is a key metric for practitioners in predicting training behavior.

\begin{figure}[t]
    \centering
    \includegraphics[width=0.9\linewidth]{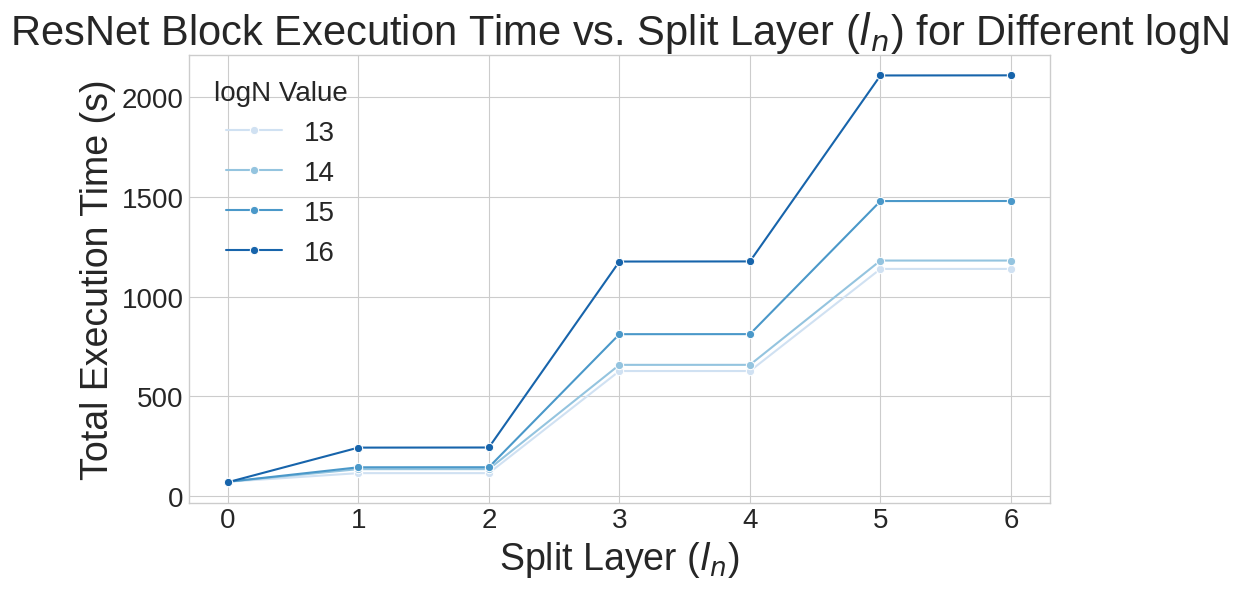}
    \caption{Total execution time of the ResNet block as a function of the split layer ($l_n$) for various CKKS ring sizes.}
    \label{fig:resnet_logN_split}
\end{figure}

\descr{Time Latency Analysis.}
We measured the end-to-end time latency for one-sample processing, which includes the forward pass, backpropagation, and weight update. Figure~\ref{fig:resnet_logN_split} presents the results for a ResNet block at $\log N=13$ on 40 cores, varying the split index. We observe that the scalability trends hold for large models, though ResNet incurs a higher execution time.

The execution time of the ResNet block stems directly from its architectural complexity, particularly the high channel depth (e.g., 64 input and 64 output channels), which demands orders of magnitude more HE operations than models like LeNet. This latency is dominated by the homomorphic \texttt{Conv2D} layers, which dramatically increase runtime. This highlights an important deployment consideration: balancing the security gains of encrypting computationally intensive layers with the associated impact on latency.


\subsubsection{Evaluation of Our Estimator}\label{sec:estimatorEval}
We conducted experiments to validate the efficacy of our estimator function (see Section~\ref{sec:estimator}). The primary success criterion is its ability to accurately predict real training times for both server and client components. Such accurate estimations are crucial for the estimator to effectively optimize training configurations under various user-defined constraints. 

\revised{Table~\ref{tab:estimator_error_splits} summarize the estimator’s quantitative fidelity across architectures, showing uniformly low deviation between predicted and actual runtimes. Overall, all models achieve less than $4.7\%$ 
error, confirming that the estimator accurately captures the scaling behavior across 
split configurations. We note that for ResNet, we implement a single block, nevertheless, the estimator can still be applied to this block. In Figure~\ref{fig:MLPEstimated} in Appendix~\ref{sec:supplementaryFigures}, we also illustrate how the estimator’s predictions follow the real MLP execution profile with near-linear correspondence across split indices. Finally, Figure~\ref{fig:LeNetEstimate}in Appendix~\ref{sec:supplementaryFigures} confirms that the estimator generalizes well to convolutional structures, preserving estimator accuracy even under deeper layer partitions.}

\begin{table}[t]
\centering
\small
\begin{tabular}{lccccc}
\toprule
\revised{\textbf{Model}} & \revised{\textbf{$l_n$=1}} & \revised{\textbf{$l_n$=3}} & \revised{\textbf{$l_n$=5}} & \revised{\textbf{$l_n$=7}} & \revised{\textbf{$l_n$=9}} \\
\midrule
\revised{MLP}        & \revised{2.6\%} & \revised{3.2\%} & \revised{3.4\%} & \revised{NA} & \revised{NA} \\
\revised{LeNet}      & \revised{2.7\%} & \revised{3.9\%} & \revised{4.7\%} & \revised{4.7\%} & \revised{4.3\%} \\
\revised{PTB-XL CNN} & \revised{2.7\%} & \revised{2.7\%} & \revised{4.1\%} & \revised{NA} & \revised{NA} \\
\midrule
\revised{ResNet} & \revised{1.7\%} & \revised{NA} & \revised{NA} & \revised{NA} & \revised{NA} \\
\bottomrule
\end{tabular}
\caption{
\revised{Estimator validation across split indices ($l_n$) for 40-core configuration.}
\revised{Estimation error is defined as 
$\frac{|\hat{t} - t_{\mathrm{real}}|}{t_{\mathrm{real}}}$,
representing the relative deviation between estimated and measured runtimes.}
\revised{For models without benchmark data at certain $l_n$ values, the same deviation trend
was propagated from available points. NA stands for "not applicable" where there is no split index existing for that network. }
}
\label{tab:estimator_error_splits}
\vspace{-1em}
\end{table}

\begin{table*}[t]
\centering
\small
\begin{tabular}{l l l l c l}
\toprule
 & \textbf{Dataset} & \textbf{Method} & \textbf{Model} & \textbf{Time (hours)} & \textbf{Notes} \\
\midrule
\multirow{4}{*}{Training} 
 & MNIST & Nandakumar et al. ~\cite{Nandakumar_Ratha_Pankanti_Halevi_2019} & \textbf{MLP} & $\sim$667 hours/epoch &  BGV  on 30 cores \\
 & MNIST & Glyph~\cite{lou2020glyph} & \textbf{MLP} & 38.4 hours/epoch & BGV  on 24 cores \\

 & \revised{MNIST} & \revised{\sys~($l_n = 1$)} & \revised{\textbf{MLP}}  & \revised{\textbf{5.72 hours/epoch}} & \revised{CKKS  on 24 cores (SL)}\\
& \revised{MNIST} & \revised{\sys~($l_n = 1$)} & \revised{\textbf{MLP}}  & \revised{\textbf{4.56 hours/epoch}} & \revised{CKKS  on 30 cores (SL)}\\
 
 & MNIST & \sys~($l_n = 1$) & \textbf{MLP}  & \textbf{3.17 hours/epoch} & CKKS  on 40 cores (SL)\\

 & \revised{MNIST} & \revised{\sys~($l_n = 5$, fully encrypted)} & \revised{\textbf{MLP}}  & \revised{\textbf{6.83 hours/epoch}} & \revised{CKKS  on 24 cores (SL)}\\
& \revised{MNIST} & \revised{\sys~($l_n = 5$, fully encrypted) } & \revised{\textbf{MLP}}  & \revised{\textbf{5.46 hours/epoch}} & \revised{CKKS  on 30 cores (SL)}\\
 
 & MNIST & \sys~($l_n = 5$, fully encrypted) & \textbf{MLP}  & \textbf{3.79 hours/epoch} &  CKKS  on 40 cores  \\
\midrule
\multirow{3}{*}{Training} 
 & PTB-XL & Khan et al.~\cite{khan2023love} & \textbf{PTB-XL CNN} & 20.15 hours/epoch &  CKKS  on 6 cores (SL) \\

& \revised{PTB-XL} & \revised{\sys~($l_n = 1$)} & \revised{\textbf{PTB-XL CNN}}  & \revised{\textbf{5.28 hours/epoch}} &  \revised{CKKS  on 6 cores (SL)} \\

 & \revised{PTB-XL} & \revised{\sys~($l_n = 5$, fully encrypted)} & \revised{\textbf{PTB-XL CNN}} & \revised{\textbf{18.93 hours/epoch}} & \revised{CKKS  on 6 cores} \\
 
 & PTB-XL & \sys~($l_n = 1$) & \textbf{PTB-XL CNN}  & \textbf{0.733 hours/epoch} &  CKKS  on 40 cores (SL) \\
 & PTB-XL & \sys~($l_n = 5$, fully encrypted) & \textbf{PTB-XL CNN}  & \textbf{2.63 hours/epoch} &  CKKS  on 40 cores \\
\midrule
\multirow{2}{*}{Inference} 
 & CIFAR-10 & Lee et al.~\cite{LeePrivacyPreservingDL} & \textbf{ResNet} &  2.94 hours/image &  CKKS  on 112 cores \\
 & CIFAR-10 & \sys & \textbf{ResNet} & \textbf{22.35 minutes/image} &  CKKS  on 40 cores (SL) \\
\bottomrule
\end{tabular}

\caption{Comparison of Training Times: Prior Work vs. \sys. All listed times are for fully encrypted training unless specified as Split Learning (SL). For fair comparison, fully encrypted \sys runtimes are obtained by configuring all layers as encrypted server layers.}
\label{tab:comparison_updated}
\end{table*}

\subsection{Comparison with Prior Work}
\label{comparison_with_prior_work}

A quantitative comparison of \sys with existing privacy-preserving SL solutions is challenging for several reasons: (i) \sys employs a novel approach, and, to the best of our knowledge, no prior work has explored the inverted traditional SL setting of \sys; (ii) we were unable to identify any fully-encrypted SL work that provides publicly available implementations for reproducibility. \revised{Consequently, we compare our framework with HE-based training methods by configuring all layers as encrypted in \sys.} 

We contrast \sys with fully encrypted training~\cite{lou2020glyph,Nandakumar_Ratha_Pankanti_Halevi_2019} and SL with HE on PTB-XL~\cite{khan2023love}, and for large models, we base our comparison on the work of Lee et al.~\cite{LeePrivacyPreservingDL}, which explores deep learning with HE. \revisedTwo{We note that Khan et al.~\cite{khan2023love} perform only a single convolutional layer under HE on the server side.} Table~\ref{tab:comparison_updated} presents our comparison results. We observe that \sys sets a new, higher standard for performance in private ML. \revised{For fully encrypted training on MNIST under the same core budget, \sys at 30 cores completes in 5.46 hours per epoch—approximately $122\times$ faster than Nandakumar et al.’s 30-core result ($\sim$667 hours/epoch) and about $7\times$ faster than the Glyph framework’s 38.4 hours/epoch result on a comparable 24-core setup.}\revisedTwo{To attribute these gains correctly, we benchmarked the underlying schemes (CKKS vs. BGV) on Lattigo. We found that CKKS yields at most a $1.11\times$ speedup for specific primitives (e.g., ciphertext-plaintext multiplication) and otherwise performs comparably to or slower than BGV and they are close on average in terms of corresponding operations latency. This confirms that \sys's substantial advantage derives from architectural optimization rather than the cryptosystem, holding firm even under worst-case primitive assumptions.} \revised{In the SL setting with matched cores, \sys at $l_n=1$ on 30 cores completes in 4.56 hours/epoch, giving a $\sim$146$\times$ speedup over Nandakumar et al.’s 30-core baseline ($\sim$667 hours/epoch).} \revised{On PTB-XL with 6-core server settings, \sys at $l_n=1$ completes in 5.28 hours/epoch versus 20.15 hours/epoch for Khan et al., a $\sim$3.8$\times$ speedup; with all layers encrypted on 6 cores, \sys is slightly faster than Khan et al. (18.93 vs. 20.15 hours/epoch).} This speedup is especially crucial as it underscores the practical advantage of the SL with HE paradigm; it enables clients to securely offload the bulk of their data storage and computational workload to a powerful server, all while maintaining confidentiality and achieving feasible training times. Finally, compared to Lee et al.’s state-of-the-art ResNet-20 inference~\cite{LeePrivacyPreservingDL}, our extrapolated results achieve 22.35 minutes per image—an $\sim$8× speedup.

\vspace{-0.25em}

\subsection{Discussion}
\label{results and analysis}
Several key insights emerge from our results: (i) the impact on model accuracy is minimal. Our findings in Section~\ref{sec:accuracy} confirm that the combination of degree-3 polynomial activation approximations and the inherent noise from the CKKS scheme results in a negligible accuracy degradation of less than 1\% across all tested models. This is a critical outcome, as it validates that the strong privacy guarantees offered by \sys do not come at the cost of the model's utility, making it a viable solution for real-world applications, (ii) The performance of \sys is dictated by a clear set of trade-offs (see Section~\ref{sec:model_time_latency}). Latency is primarily determined by the number of encrypted layers, hardware capacity, and cryptographic parameters. Core-count scalability analysis shows near-linear speedup, making the framework well-suited to modern multi-core servers. However, this efficiency must be contextualized by the overhead of HE itself. As shown in our \emph{split layer index} and $\log N$ experiments, both deepening the encrypted portion of the model and increasing the security parameters lead to substantial increases in runtime. This underscores the necessity of our estimator tool, which is designed to navigate these trade-offs and identify an optimal configuration for a given set of constraints. (iii) Our analysis of large models shows that convolutional layers are the main performance bottleneck. In ResNet, high channel depth makes Conv2D layers vastly more expensive under HE than layers in simpler models like LeNet, underscoring the importance of carefully choosing the split point. (iv) Benchmark results show that \sys delivers a substantial performance leap. As Table~\ref{tab:comparison_updated} illustrates, it is up to $210\times$ faster than foundational HE training frameworks and consistently outperforms recent specialized competitors. The advantage is especially clear on the PTB-XL dataset, where \sys’s SL implementation is orders of magnitude faster than Khan et al. Overall, these results establish \sys as competitive and practical.

\revised{Finally, we note that while recent works such as BOLT~\cite{Pang2024BOLT}, BumbleBee~\cite{Lu2025BumbleBee}, and NEXUS~\cite{Zhang2025NEXUS} have paved the way for secure inference on transformers—where a service provider holds \textit{static} model weights and only the user inputs are encrypted—our contribution is orthogonal: we enable server-side training with \textit{dynamic encrypted} model parameters with \textit{backpropagation} under SL. 
Rather than hard-coding specific neural networks, we implement layers as composable modules that interoperate under any split index and dimension configuration. Our packing strategies are designed to remain efficient even as layer sets or split points vary during training. In contrast, \cite{Pang2024BOLT, Lu2025BumbleBee, Zhang2025NEXUS} propose optimizations tailored primarily to secure inference of transformer architectures, relying on techniques such as attention-specific packing layouts, low-rotation matrix multiplication schemes, and efficient approximations of softmax, assuming fixed weights and forward-only execution. In particular, the matrix-multiplication methods in \cite{Pang2024BOLT, Lu2025BumbleBee, Zhang2025NEXUS} are tailored for fixed-weight, inference-only settings and cannot be directly applied to CURE’s encrypted-training scenario: BOLT’s optimizations rely on the server being able to freely access and pre-arrange its model parameters in plaintext form to minimize rotation and packing costs during computation; BumbleBee’s Oblivious Linear Transformation (OLT) operates over secret-shared inputs in a two-party computation framework which contradicts with CURE's system/threat model; and NEXUS’s optimization is built on having fixed parameters that allow pre-computation and reuse of encrypted operations across multiple inputs. Since CURE maintains and updates (during backpropagation) all server-side parameters as ciphertexts under the client’s key during training, none of these methods are compatible with its homomorphic, single-key, SL design.}

\revised{\paragraph{Future Work}
\revisedTwo{While this work focuses on the single-client setting, an interesting future direction is to extend \sys to multi-party SL, enabling multiple clients to jointly train a model under semi-honest or malicious settings. However, we note that this necessitates distinct cryptographic solutions, such as distributed key management and the aggregation of encrypted updates. Multi-party environments introduce non-trivial challenges—such as data integrity and peer-wise privacy—that are not present in the single-client architecture of \sys.} One solution would be to integrate multiparty HE~\cite{mouchet2021multiparty} for secure key generation and aggregation, eliminating reliance on a trusted party as in prior private federated learning systems~\cite{poseidon}. Moreover, unlike the single-client setting—where all updates originate from a single honest party—the multi-client scenario introduces risks of malicious or inconsistent client behavior, such as submitting malformed or adversarial gradients. Therefore, robust and verifiable mechanisms, including encrypted gradient validation and zero-knowledge proofs, are critical to ensure correctness and integrity. The estimator may also be generalized to manage dynamic cut-layer placement, heterogeneous computational capacities, and communication scheduling across clients. Finally, formalizing the multi-party security model and assessing \sys in cross-silo settings (e.g., multi-hospital collaborations) would further validate its scalability, robustness, and privacy guarantees.}


\section{Conclusion}\label{sec:conclusion}
We introduced \sys, an efficient privacy-preserving split-learning framework for a single-client, single-server setting that encrypts only server-side parameters. This design securely offloads storage and computation while preserving label confidentiality and, optionally, data privacy. 
By relying on our proposed packing strategies, \sys further enhances performance and outperforms fully encrypted training methods while achieving accuracy levels on par with both baseline and fully encrypted approaches. 

\bibliographystyle{abbrv}
\bibliography{bibfile}

\appendix

\section{\sys's Initialization Algorithm}\label{sec:supplementaryAlgo}
We provide the initialization of \sys that involves the generation and encryption of weight matrices described in Section~\ref{System Overview} in Algorithm~\ref{Innitialization}. 

\begin{algorithm}[h]
\caption{Initialization}
\begin{algorithmic}[1]
    \Function{Initialization}{}
        \If{client}
        \State $L_c \leftarrow{[l_{n+1}, \ldots, l_{n+k}]}$
        \Comment{Client-side layers}
        \State $W_c \gets \text{GenRandomWeights}(L_c)$
        \State $(\textsf{PK}_c, \textsf{SK}_c) \gets \text{KeyGen}(1^{\lambda})$
        \State \text{Send} $\textsf{PK}_c$ \text{to server}
        
        \ElsIf{server}
        \State $L_s \leftarrow{[l_1, l_2, \ldots, l_n]}$
        \Comment{Server-side layers}
        \State $W_s \gets \text{GenRandomWeights}(L_s)$
        \State $\mathbf{W}_s \gets \text{Enc}_{\textsf{PK}_c}(W_s)$     
\EndIf
    \EndFunction
\end{algorithmic}
\label{Innitialization}
\end{algorithm}
\section{Illustration of Packing Principles}\label{sec:supplementaryPacking}
The following example outlines how we apply packing principle through a toy example. Here \( d_{ij} \) represents the entries of an arbitrary matrix, \(\beta\) is a scalar multiplicative, and underscores represent garbage values. Consider $D$ a $3 \times 3$ data matrix:
\vspace{0.3em} 
\begin{center}
\[D = \begin{bmatrix}
\mathbf{d_{00}} & \mathbf{d_{01}} & \mathbf{d_{02}} \\
\mathbf{d_{10}} & \mathbf{d_{11}} & \mathbf{d_{12}} \\
\mathbf{d_{20}} & \mathbf{d_{21}} & \mathbf{d_{22}}
\end{bmatrix}\]
\end{center}
\vspace{0.3em} 
Instead of performing a scalar multiplication in HE separately for each entry of the matrix as:
\vspace{0.3em} 
\[
\begin{array}{c}
\beta\cdot \begin{bmatrix} \mathbf{d_{00}} & \_ & \_ & \_ & \_ & \dots & \_ \end{bmatrix} \\
\beta \cdot \begin{bmatrix} \mathbf{d_{01}} & \_ & \_ & \_ & \_ & \dots & \_ \end{bmatrix} \\
\vdots \\
\beta \cdot \begin{bmatrix} \mathbf{d_{22}} & \_ & \_ & \_ & \_ & \dots & \_ \end{bmatrix} \\
\end{array}
\]
\vspace{0.3em} 
We pack and augment the data properly with respect to the operation to utilize the computational resources more efficiently. The packing is done as follows:
\vspace{0.3em} 
\[
\beta \cdot \begin{bmatrix} \mathbf{d_{00}} & \mathbf{d_{01}} & \dots & \mathbf{d_{22}} & \_ & \dots & \_ \end{bmatrix}
\]
\vspace{0.3em} 
By restructuring the data in this manner, we fill the RLWE vector with meaningful data and pad it with zeros when necessary, as described in Section~\ref{sec:he_linear}. 

\section{One-Level Operations in Matrix Computation}\label{sec:supplementaryOneLevel}

In one-level batch multiplication, we pack weight matrices as batches of columns, enabling matrix-vector multiplication. Optimizing the second layer's matrix initialization is crucial since packing efficiency depends on the dimensions of the first two layers and the number of slots in the RLWE vector. This process is complex and influenced by the dataset, HE scheme parameters, and security and computational limits. While packing the encrypted weight matrix for batch multiplication can be costly (as discussed in Section~\ref{comparison_with_prior_work}), it maximizes efficiency. The latency from batch multiplication is compensated by the performance boost from better weight matrix packing.

In contrast, for one-level scalar multiplication, there is no need for such preprocessing on the components of the network except for the encoding and encryption of the elements. However, since one-level scalar multiplication cannot utilize packing as efficiently as one-level batch multiplication, there is a possibility of higher demand for memory and computation in some cases. Therefore, it is important to carefully decide which one-level operation to use, considering the trade-offs.

Here, we provide a toy example to demonstrate the necessary operations to calculate the multiplication of a $4\times4$ matrix with a $4$-dimensional vector:
\vspace{0.5em}
\[
\scriptsize
\begin{aligned}
&\begin{bmatrix}
    \mathbf{v_{11}} & \mathbf{v_{12}} & \mathbf{v_{13}} & \mathbf{v_{14}} \\
    \mathbf{v_{21}} & \mathbf{v_{22}} & \mathbf{v_{23}} & \mathbf{v_{24}} \\
    \mathbf{v_{31}} & \mathbf{v_{32}} & \mathbf{v_{33}} & \mathbf{v_{34}} \\
    \mathbf{v_{41}} & \mathbf{v_{42}} & \mathbf{v_{43}} & \mathbf{v_{44}} 
    \end{bmatrix}
    \times
    \begin{bmatrix}
    w_{1}  \\
    w_{2}  \\
    w_{3}  \\
    w_{4}  
    \end{bmatrix}
    =
    \begin{bmatrix}
    \mathbf{v_{11} w_1 + v_{12} w_2 + v_{13} w_3 + v_{14} w_4} \\
    \mathbf{v_{21} w_1 + v_{22} w_2 + v_{23} w_3 + v_{24} w_4} \\
    \mathbf{v_{31} w_1 + v_{32} w_2 + v_{33} w_3 + v_{34} w_4} \\
    \mathbf{v_{41} w_1 + v_{42} w_2 + v_{43} w_3 + v_{44} w_4}
    \end{bmatrix}
\end{aligned}
\]

Our one-level batch multiplication is represented as:
\vspace{0.5em} 
\[
\scriptsize
= 
    \begin{bmatrix}
    w_1 \\
    w_1 \\
    w_1 \\
    w_1 
    \end{bmatrix}
    \odot
    \begin{bmatrix}
    \mathbf{v_{11}} \\
    \mathbf{v_{21}} \\
    \mathbf{v_{31}} \\
    \mathbf{v_{41}} 
    \end{bmatrix}
    +
    \begin{bmatrix}
    w_2 \\
    w_2 \\
    w_2 \\
    w_2 
    \end{bmatrix}
    \odot
    \begin{bmatrix}
    \mathbf{v_{12}} \\
    \mathbf{v_{22}} \\
    \mathbf{v_{32}} \\
    \mathbf{v_{42}} 
    \end{bmatrix}
    +
    \begin{bmatrix}
    w_3 \\
    w_3 \\
    w_3 \\
    w_3 
    \end{bmatrix}
    \odot
    \begin{bmatrix}
    \mathbf{v_{13}} \\
    \mathbf{v_{23}} \\
    \mathbf{v_{33}} \\
    \mathbf{v_{43}} 
    \end{bmatrix}
    +
    \begin{bmatrix}
    w_4 \\
    w_4 \\
    w_4 \\
    w_4 
    \end{bmatrix}
    \odot
    \begin{bmatrix}
    \mathbf{v_{14}} \\
    \mathbf{v_{24}} \\
    \mathbf{v_{34}} \\
    \mathbf{v_{44}} 
    \end{bmatrix}
\]
\vspace{0.3em} 
And our one-level scalar multiplication is represented as:
\vspace{0.3em} 
\[
\scriptsize
= 
    w_1
    \otimes
    \begin{bmatrix}
    \mathbf{v_{11}} \\
    \mathbf{v_{21}} \\
    \mathbf{v_{31}} \\
    \mathbf{v_{41}} 
    \end{bmatrix}
    +
    w_2
    \otimes
    \begin{bmatrix}
    \mathbf{v_{12}} \\
    \mathbf{v_{22}} \\
    \mathbf{v_{32}} \\
    \mathbf{v_{42}} 
    \end{bmatrix}
    +
   w_3
    \otimes
    \begin{bmatrix}
    \mathbf{v_{13}} \\
    \mathbf{v_{23}} \\
    \mathbf{v_{33}} \\
    \mathbf{v_{43}} 
    \end{bmatrix}
    +
   w_4
    \otimes
    \begin{bmatrix}
    \mathbf{v_{14}} \\
    \mathbf{v_{24}} \\
    \mathbf{v_{34}} \\
    \mathbf{v_{44}} 
    \end{bmatrix}
\]
\vspace{0.3em} 
Notice that the column-wise multiplication can be done in a batch or scalar fashion. In other words, we can write the multiplication of a column as scalar multiplication of $w_i$ or element-wise vector multiplication with repeated elements of $w_i$ as an RLWE-packed vector. When the ciphertext batch size is larger than the column size (e.g., two columns fit in one RLWE vector), we can utilize packing more efficiently for the one-level batch multiplication operation, as follows:

\begin{center}
\[
\begin{bmatrix}
w_1 \\
w_1 \\
w_1 \\
w_1 \\
w_2 \\
w_2 \\
w_2 \\
w_2 
\end{bmatrix}
\odot
\begin{bmatrix}
\mathbf{v_{11}} \\
\mathbf{v_{21}} \\
\mathbf{v_{31}} \\
\mathbf{v_{41}} \\
\mathbf{v_{12}} \\
\mathbf{v_{22}} \\
\mathbf{v_{32}} \\
\mathbf{v_{42}} 
\end{bmatrix}
+
\begin{bmatrix}
w_3 \\
w_3 \\
w_3 \\
w_3 \\
w_4 \\
w_4 \\
w_4 \\
w_4 
\end{bmatrix}
\odot
\begin{bmatrix}
\mathbf{v_{13}} \\
\mathbf{v_{23}} \\
\mathbf{v_{33}} \\
\mathbf{v_{43}} \\
\mathbf{v_{14}} \\
\mathbf{v_{24}} \\
\mathbf{v_{34}} \\
\mathbf{v_{44}} 
\end{bmatrix}
=
\begin{bmatrix}
\mathbf{v_{11} * w_1 + v_{13} * w_3}\\
\mathbf{v_{21} * w_1 + v_{23} * w_3}\\
\mathbf{v_{31} * w_1 + v_{33} * w_3}\\
\mathbf{v_{41} * w_1 + v_{43} * w_3}\\
\mathbf{v_{12} * w_2 + v_{14} * w_4}\\
\mathbf{v_{22} * w_2 + v_{24} * w_4}\\
\mathbf{v_{32} * w_2 + v_{34} * w_4}\\
\mathbf{v_{42} * w_2 + v_{44} * w_4}
\end{bmatrix}
\]
\end{center}
Upon receiving the ciphertext, the client decrypts it and calculates:

\[
\scriptsize
\begin{bmatrix}
v_{11} w_1 + v_{13} w_3 \\
v_{21} w_1 + v_{23} w_3 \\
v_{31} w_1 + v_{33} w_3 \\
v_{41} w_1 + v_{43} w_3 \\
\end{bmatrix}
+
\begin{bmatrix}
v_{12} w_2 + v_{14} w_4 \\
v_{22} w_2 + v_{24} w_4 \\
v_{32} w_2 + v_{34} w_4 \\
v_{42} w_2 + v_{44} w_4
\end{bmatrix}
=
\begin{bmatrix}
v_{11} w_1 + v_{12} w_2 + v_{13} w_3 + v_{14} w_4 \\
v_{21} w_1 + v_{22} w_2 + v_{23} w_3 + v_{24} w_4 \\
v_{31} w_1 + v_{32} w_2 + v_{33} w_3 + v_{34} w_4 \\
v_{41} w_1 + v_{42} w_2 + v_{43} w_3 + v_{44} w_4 
\end{bmatrix}
\normalsize
\]

\section{Illustration $n$-encrypted Layer Operations in Matrix Multiplication}\label{sec:supplementaryNLevel}
Here, we provide a toy example that illustrates the packing and matrix multiplication scheme for $n$-encrypted layer networks described in Section~\ref{sec:he_linear}. Consider a matrix $3\times 4$ matrix $B$ with the following entries:

\[
\mathbf{B} = \begin{bmatrix}
w_{00} & w_{01} & w_{02} & w_{03} & w_{04} \\
w_{10} & w_{11} & w_{12} & w_{13} & w_{14} \\
w_{20} & w_{21} & w_{22} & w_{23} & w_{24} \\
\end{bmatrix}
\]

First, we pad our matrix to achieve an efficient homomorphic dot product with optimized rotations.

\[
\begin{bmatrix}
\begin{array}{c}
w_{00} \\ w_{10} \\ w_{20} \\ 0
\end{array}
\quad
\begin{array}{c}
w_{01} \\ w_{11} \\ w_{21} \\ 0
\end{array}
\quad \cdots \quad
\begin{array}{c}
w_{04} \\ w_{14} \\ w_{24} \\ 0
\end{array}
\end{bmatrix}
\]

By concatenating and marking the entries, we achieve the placement of columns to the RLWE vectors for the dot product.

\[
\begin{bmatrix} \underline{w_{00}} \\ w_{10} \\ w_{20} \\ 0 \\ \underline{w_{01}} \\ w_{11} \\ w_{21} \\ 0 \end{bmatrix} 
\begin{bmatrix} \underline{w_{02}} \\ w_{12} \\ w_{22} \\ 0 \\ \underline{w_{03}} \\ w_{13} \\ w_{23} \\ 0 \end{bmatrix} 
\begin{bmatrix} \underline{w_{04}} \\ w_{14} \\ w_{24} \\ 0 \\ 0 \\ 0 \\ 0 \\ 0 \end{bmatrix} 
\]

After preparing the second vector, we process the rows of the first matrix by padding them to the nearest power of two and repeating the rows as necessary. We then calculate the homomorphic dot product for each column and extract the previously marked data. It is important to note that this marking operation serves as an abstraction for explanatory purposes.

\[
\mathbf{A} =\begin{bmatrix}
v_{00} & v_{11} & v_{02} \\
v_{10} & v_{11} & v_{12} \\
v_{20} & v_{21} & v_{22} \\
\end{bmatrix}
\]

To prepare the first row for the homomorphic dot product calculation, we arrange the elements as $[ v_{00}, v_{10}, v_{20}, 0, v_{00}, v_{00}, v_{00}, 0, ...]$ as a column vector.
Next, we perform element-wise multiplication of this vector with the columns obtained from matrix \(\mathbf{B}\). Subsequently, we rotate the resulting vector by powers of two until we cover all slots. After each rotation, we perform an element-wise addition with the previously accumulated vector. This process ultimately yields the dot product of the initial matrices' row-column pairs homomorphically, enabling efficient computation of the matrix-matrix product. Importantly, all operations from this stage onward are executed homomorphically.

\[
\begin{bmatrix}
\mathbf{w_{00}} \\
\mathbf{w_{10}} \\
\mathbf{w_{20}} \\
\mathbf{0} \\
\mathbf{w_{01}} \\
\mathbf{w_{11}} \\
\mathbf{w_{21}} \\
\mathbf{0} 
\end{bmatrix}
\odot
\begin{bmatrix}
\mathbf{v_{00}} \\
\mathbf{v_{10}} \\
\mathbf{v_{20}} \\
\mathbf{0} \\
\mathbf{v_{01}} \\
\mathbf{v_{10}} \\
\mathbf{v_{21}} \\
\mathbf{0} 
\end{bmatrix}
=
\begin{bmatrix}
\mathbf{v_{00}} * \mathbf{w_{00}} \\
\mathbf{v_{10}} * \mathbf{w_{10}}  \\
\mathbf{v_{20}} * \mathbf{w_{20}}  \\
\mathbf{0} * \mathbf{0}  \\
\mathbf{v_{01}} * \mathbf{w_{02}} \\
\mathbf{v_{10}} * \mathbf{w_{11}}  \\
\mathbf{v_{21}} * \mathbf{w_{21}} \\
\mathbf{0} * \mathbf{0} 
\end{bmatrix}
\]

Next, we rotate and add the result to itself $\lceil\log n\rceil -1$
times, ultimately achieving the desired outcome:

\[
\scriptsize
    \begin{bmatrix}
    \mathbf{v_{00}} * \mathbf{w_{00}} \\
    \mathbf{v_{10}} * \mathbf{w_{10}}  \\
    \mathbf{v_{20}} * \mathbf{w_{20}}  \\
    \mathbf{0} * \mathbf{0}  \\
    \mathbf{v_{01}} * \mathbf{w_{02}} \\
    \mathbf{v_{10}} * \mathbf{w_{11}}  \\
    \mathbf{v_{21}} * \mathbf{w_{21}} \\
    \mathbf{0} * \mathbf{0} 
    \end{bmatrix}
    +
    \begin{bmatrix}
    \mathbf{v_{10}} * \mathbf{w_{10}}  \\
    \mathbf{v_{20}} * \mathbf{w_{20}}  \\
    \mathbf{0} * \mathbf{0}  \\
    \mathbf{v_{01}} * \mathbf{w_{02}} \\
    \mathbf{v_{10}} * \mathbf{w_{11}}  \\
    \mathbf{v_{21}} * \mathbf{w_{21}} \\
    \mathbf{0} * \mathbf{0} \\
    \mathbf{v_{00}} * \mathbf{w_{00}} 
    \end{bmatrix}
    =
    \begin{bmatrix}
    \mathbf{v_{00}} * \mathbf{w_{00}} +  \mathbf{v_{10}} * \mathbf{w_{10}}  \\
    \mathbf{v_{10}} * \mathbf{w_{10}}  +\mathbf{v_{20}} * \mathbf{w_{20}}  \\
    \mathbf{v_{20}} * \mathbf{w_{20}}  +\mathbf{0} * \mathbf{0}  \\
    \mathbf{0} * \mathbf{0}  +\mathbf{v_{01}} * \mathbf{w_{02}} \\
    \mathbf{v_{01}} * \mathbf{w_{02}} +\mathbf{v_{10}} * \mathbf{w_{11}}  \\
    \mathbf{v_{10}} * \mathbf{w_{11}}  +\mathbf{v_{21}} * \mathbf{w_{21}} \\
    \mathbf{v_{21}} * \mathbf{w_{21}} +\mathbf{0} * \mathbf{0} \\
    \mathbf{0} * \mathbf{0} +\mathbf{v_{00}} * \mathbf{w_{00}} \\
    \end{bmatrix}
\]\\

\noindent Rotation of this vector twice and addition will result in:

\[
\scriptsize
    \begin{bmatrix}
    \mathbf{v_{00}} * \mathbf{w_{00}} +  \mathbf{v_{10}} * \mathbf{w_{10}}  \\
    \mathbf{v_{10}} * \mathbf{w_{10}}  +\mathbf{v_{20}} * \mathbf{w_{20}}  \\
    \mathbf{v_{20}} * \mathbf{w_{20}}  +\mathbf{0} * \mathbf{0}  \\
    \mathbf{0} * \mathbf{0}  +\mathbf{v_{01}} * \mathbf{w_{02}} \\
    \mathbf{v_{01}} * \mathbf{w_{02}} +\mathbf{v_{10}} * \mathbf{w_{11}}  \\
    \mathbf{v_{10}} * \mathbf{w_{11}}  +\mathbf{v_{21}} * \mathbf{w_{21}} \\
    \mathbf{v_{21}} * \mathbf{w_{21}} +\mathbf{0} * \mathbf{0} \\
    \mathbf{0} * \mathbf{0} +\mathbf{v_{00}} * \mathbf{w_{00}} \\
    \end{bmatrix}
    +
    \begin{bmatrix}
\mathbf{v_{20}} *   \mathbf{w_{20}}  +  \mathbf{0} *        \mathbf{0}  \\
\mathbf{0} *        \mathbf{0}  +       \mathbf{v_{01}} *   \mathbf{w_{02}} \\
\mathbf{v_{01}} *   \mathbf{w_{02}} +   \mathbf{v_{10}} *   \mathbf{w_{11}}  \\
\mathbf{v_{10}} *   \mathbf{w_{11}}  +  \mathbf{v_{21}} *   \mathbf{w_{21}} \\
\mathbf{v_{21}} *   \mathbf{w_{21}} +   \mathbf{0} *        \mathbf{0} \\
\mathbf{0} *        \mathbf{0} +        \mathbf{v_{00}} *   \mathbf{w_{00}} \\
\mathbf{v_{00}} *   \mathbf{w_{00}} +   \mathbf{v_{10}} *   \mathbf{w_{10}}  \\
\mathbf{v_{10}} *   \mathbf{w_{10}}  +  \mathbf{v_{20}} *   \mathbf{w_{20}}  
\end{bmatrix}
\]

\[
\scriptsize
=
\begin{bmatrix}
\mathbf{v_{00}} * \mathbf{w_{00}} +  \mathbf{v_{10}} * \mathbf{w_{10}}  + \mathbf{v_{20}} *   \mathbf{w_{20}}  +  \mathbf{0} *        \mathbf{0}  \\
\mathbf{v_{10}} * \mathbf{w_{10}}  +\mathbf{v_{20}} * \mathbf{w_{20}}  +\mathbf{0} *        \mathbf{0}  +       \mathbf{v_{01}} *   \mathbf{w_{02}} \\
\mathbf{v_{20}} * \mathbf{w_{20}}  +\mathbf{0} * \mathbf{0}  +\mathbf{v_{01}} *   \mathbf{w_{02}} +   \mathbf{v_{10}} *   \mathbf{w_{11}}  \\
\mathbf{0} * \mathbf{0}  +\mathbf{v_{01}} * \mathbf{w_{02}} +\mathbf{v_{10}} *   \mathbf{w_{11}}  +  \mathbf{v_{21}} *   \mathbf{w_{21}} \\
\mathbf{v_{01}} * \mathbf{w_{02}} +\mathbf{v_{10}} * \mathbf{w_{11}}  +\mathbf{v_{21}} *   \mathbf{w_{21}} +   \mathbf{0} *        \mathbf{0} \\
\mathbf{v_{10}} * \mathbf{w_{11}}  +\mathbf{v_{21}} * \mathbf{w_{21}} +\mathbf{0} *        \mathbf{0} +        \mathbf{v_{00}} *   \mathbf{w_{00}} \\
\mathbf{v_{21}} * \mathbf{w_{21}} +\mathbf{0} * \mathbf{0} +\mathbf{v_{00}} *   \mathbf{w_{00}} +   \mathbf{v_{10}} *   \mathbf{w_{10}}  \\
\mathbf{0} * \mathbf{0} +\mathbf{v_{00}} * \mathbf{w_{00}} +\mathbf{v_{10}} *   \mathbf{w_{10}}  +  \mathbf{v_{20}} *   \mathbf{w_{20}}  
\end{bmatrix}
\]

\vspace{1em}
Note that the first and fifth entries of the final vector represent the desired homomorphic and efficient results of the first row's first column and the first row's second column of the resulting matrix. 
By proceeding with this process for each column and row, we can obtain the matrix-matrix product. For matrices with columns that do not fit into a single RLWE vector, we perform additional summation operations on the final result, based on the initial slot-to-column length ratio calculation.
\section{Approximated Activation Functions}
\label{sec:ActivationFunction}
Due to the fully encrypted nature of the server layers, for $n$ encrypted server layers where $n > 1$, the activation functions of $n-1$ layers should also be executed under encryption. However, non-linear activation functions cannot be directly applied under encryption; only polynomial functions can be used. To address this limitation, we rely on well-known approximation techniques such as the Chebyshev interpolation method~\cite{press2007numerical} or the minimax approximation~\cite{stephens2015big} to approximate the non-linear activation functions as polynomials. This technique is also employed by numerous privacy-preserving machine learning works~\cite{poseidon,Maya2020,CryptoNets,rhode,froelicher2020scalable,cryptoDL,minionn}. It is important to note that using higher-degree polynomials may result in better approximations and thus better accuracy. However, higher-degree polynomials also lead to more HE multiplications, resulting in noise accumulation and potentially necessitating bootstrapping as each multiplication consumes one ciphertext level. For a degree $d$ polynomial, the scheme consumes $log_2(d+1)$ levels. This results in increased computational complexity and can lead to higher training latency. Therefore, careful consideration is required when selecting the function and the degree of the polynomial used for the approximation.

\paragraph{Backward Pass for Activation Functions}
Activation functions, being non-parametric, do not require weight updates. During the backward pass, the gradient with respect to the activation function's input is computed. This is achieved by multiplying the incoming gradient (with respect to the activation's output) by the derivative of the activation function itself. For our approximated polynomial activations, we directly use the derivative of the polynomial approximation to compute this gradient.

\section{Detailed Implementation Details}\label{sec:implementationAppendix}

Every NN layer used in experiments has \emph{both} a plaintext reference implementation and an HE counterpart. This lets us (i) execute \sys with end-to-end HE encryption, meaning all computations—from forward pass to backpropagation—are performed on encrypted data without any intermediate decryptions except client operations and (ii) quantify per-layer divergence by comparing HE outputs to their plaintext mirrors at each split layer index ($l_n$). For accuracy testing, we also provide a Python/PyTorch simulation harness that mirrors our Go operators by using the same fixed-point quantization and degree-3 polynomial approximation of ReLU, and injecting layer-wise noise calibrated from our measurements and Lattigo’s CKKS noise growth reports. Unless stated otherwise, each experiment is repeated at least three times, and we report the mean.
\section{Detailed Information on Datasets}\label{sec:supplementaryDatasets}
We detail the datasets used in the evaluation of \sys here. We employ: (i) the hand-written digits (MNIST) dataset~\cite{lecun-mnisthandwrittendigit-2010}  with $60,000$ images of $28 \times 28$ pixels and 10 labels, (ii) the PTB XL dataset~\cite{Wagner_PTB} (PTB-XL) with $21,837$ clinical 12-lead ECG records, 10-second recordings with $100$ Hz sampling rate, annotated with up to $71$ different diagnostic classes, and (iii) the CIFAR-10 dataset~\cite{krizhevsky2009learning} with 60,000 color images of $32 \times 32$ pixels, categorized into 10 classes.
\section{Experimental Methodology}
\label{sec:methodology}
\descr{Time latency.}
All time-latency experiments run true end-to-end HE pipelines. For every layer, we retain both plaintext and HE implementations and log the per-layer HE\(\leftrightarrow\)plaintext deviation to monitor drift.
Each layer processes the actual output of its predecessor, thereby capturing real training dynamics such as ciphertext scaling, rescaling, and relinearization.

\descr{Accuracy via calibrated noise-injection.}
\label{sec:accuracy-method}
To estimate accuracy under HE, we finetune training for additional epochs in PyTorch while injecting per-layer noise that emulates our measured HE deviation and Lattigo-reported CKKS noise growth. This is performed on Colab T4~\cite{google-colab} for fast iteration. We ensure the activation approximation and quantization match the Go/HE pipeline.

\descr{Scalability experiments.}
 We measured scalability with respect to two primary factors: the number of CPU cores and the cryptographic context parameters, specifically by varying the ring dimension, \(\log N\). For these experiments, we report the total wall-clock time for end-to-end model passes. Additionally, to provide a more granular analysis, we also benchmarked representative cryptographic primitives (e.g., $ct$\(\times\)$pt$, $ct$\(\times\)$ct$, and rotations).

 \section{Supplementary Figures and Tables}~\label{sec:supplementaryFigures}
We present here the supplementary figures and tables here that could not be included in the main manuscript due to space constraints. Discussions on these figures and tables can be found in the relevant subsections of Section~\ref{evaluation}.
 \begin{figure}[h]
    \centering
    \begin{minipage}[b]{0.48\linewidth}
        \centering
        \includegraphics[width=\linewidth]{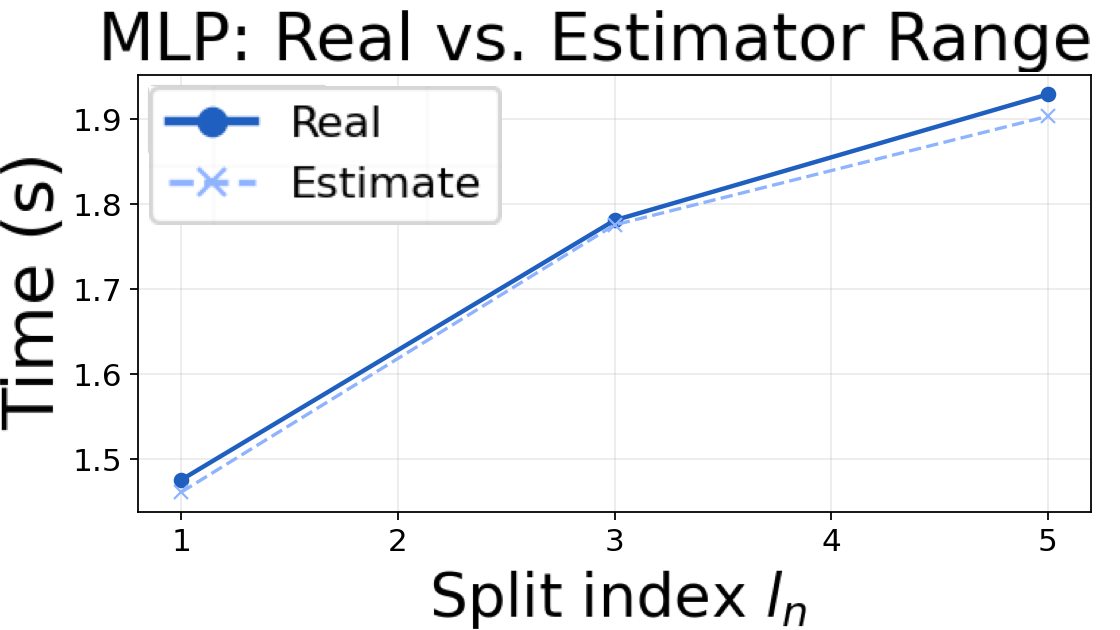}
        \vspace{-0.8em}
        \caption{\revised{MLP real and estimator times}}
        \label{fig:MLPEstimated}
    \end{minipage}
    \hfill
    \begin{minipage}[b]{0.48\linewidth}
        \centering
        \includegraphics[width=\linewidth]{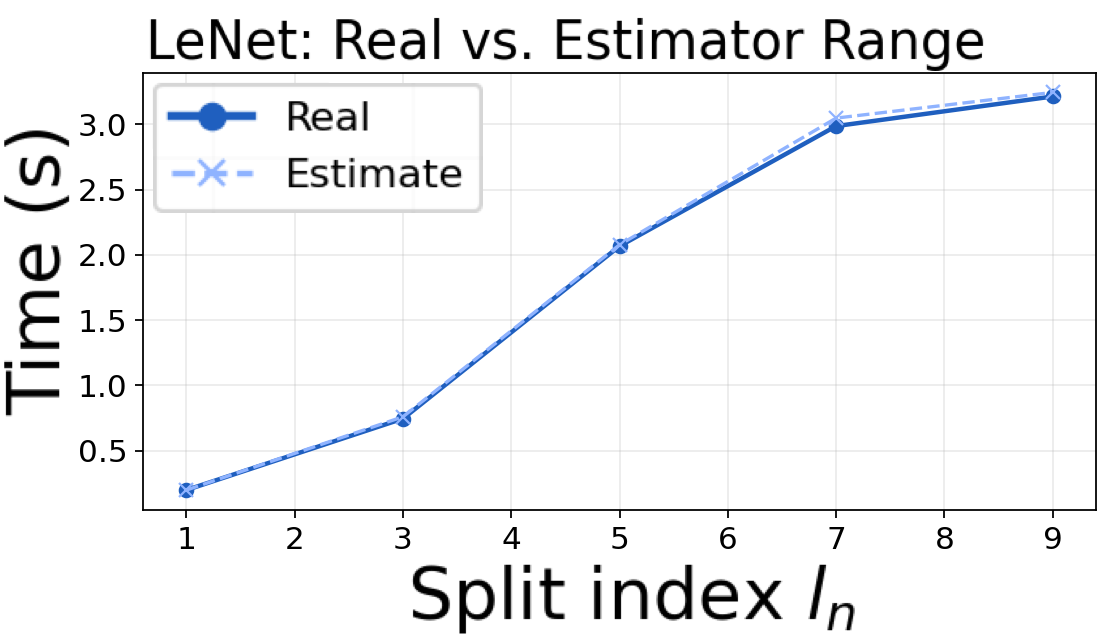}
        \vspace{-0.8em}
        \caption{\revised{LeNet real and estimator times}}
        \label{fig:LeNetEstimate}
    \end{minipage}
\end{figure}
\end{document}